# A Review of Flood Risk Assessment Frameworks and the Development of Hierarchical Structures for Risk Components

Nazgol Tabasi[1], Mohammad Fereshtehpour[2*], Bardia Roghani[3]

[1] Department of Industrial Engineering, Faculty of Engineering, Ferdowsi University of Mashhad, Mashhad, Iran, tabasi74.nazgol@gmail.com
[2*] Department of Civil and Environmental Engineering, University of Western Ontario, ON Canada, mferesht@uw.o.ca
[3] Norwegian University of Life Sciences, Institute of Civil and Environmental Engineering, Faculty of Science and Technology (REALTEK), Ås, Norway, bardia.roghani@nmbu.no

**Abstract**

Climate change and rapid urbanization have intensified the frequency and severity of flooding, resulting in substantial damage to communities and infrastructure. Existing research on flood risk addresses a wide range of dimensions, ranging from physical to managerial aspects, which adds complexity to the assessment process. This paper introduces the Integrated Risk Linkages (IRL) Framework to provide a systematic approach to flood risk assessment. The IRL Framework defines risk as the intersection of hazard and vulnerability, where vulnerability is shaped by exposure and susceptibility. Resilience, including coping and adaptive capacities, serves as a counterbalance to vulnerability, offering pathways to mitigate flood impacts. Guided by the IRL framework, this study conducts a comprehensive review of the literature to identify and organize a detailed set of 91 criteria and sub-criteria into three hierarchical structures: hazard, susceptibility, and resilience. Furthermore, the paper evaluates existing flood risk assessment methods, emphasizing their characteristics and practical applicability. The IRL framework presented in this study offers essential insights for navigating the complexities of flood risk management, serving as a valuable reference for researchers, policymakers, and practitioners. Its flexibility empowers users to adapt the framework by utilizing specific components or its entire hierarchical structure, depending on data availability and research objectives, thereby enhancing its applicability across diverse contexts.

**Keywords:** Flood risk, Hazard, Vulnerability, Exposure, Resilience, Risk mitigation.

**Article Highlights:**
- This study develops the Integrated Risk Linkages (IRL) Framework to provide a systematic approach to flood risk assessment.
- A comprehensive review of existing literature is conducted, organizing 91 criteria into hierarchical structures for hazard, susceptibility, and resilience assessments.
- The paper addresses challenges in data quality and integration by promoting flexibility and adaptability in applying the IRL framework along with the proposed hierarchical structures across diverse contexts.

# 1. Introduction

Natural hazards cause severe damages to the environment and humans. They may cause loss of life, destruction of infrastructures and property, and disruption of economic and social activities [1]. The impacts of these events appear to be escalating, primarily due to their heightened intensity and the greater value of properties exposed to them. Among various natural hazards, flood hazards are particularly notorious for their devastating effects, and their occurrence has become increasingly frequent in recent decades [1–5]. Global statistics reveal that both flood damages and the number of people affected by floods have experienced significant increases and have remained consistently high [2]. Figure 1 illustrates that floods were among the most frequent catastrophic events, with 170 occurrences in 2023 and an average of 164 occurrences in the years prior[6].

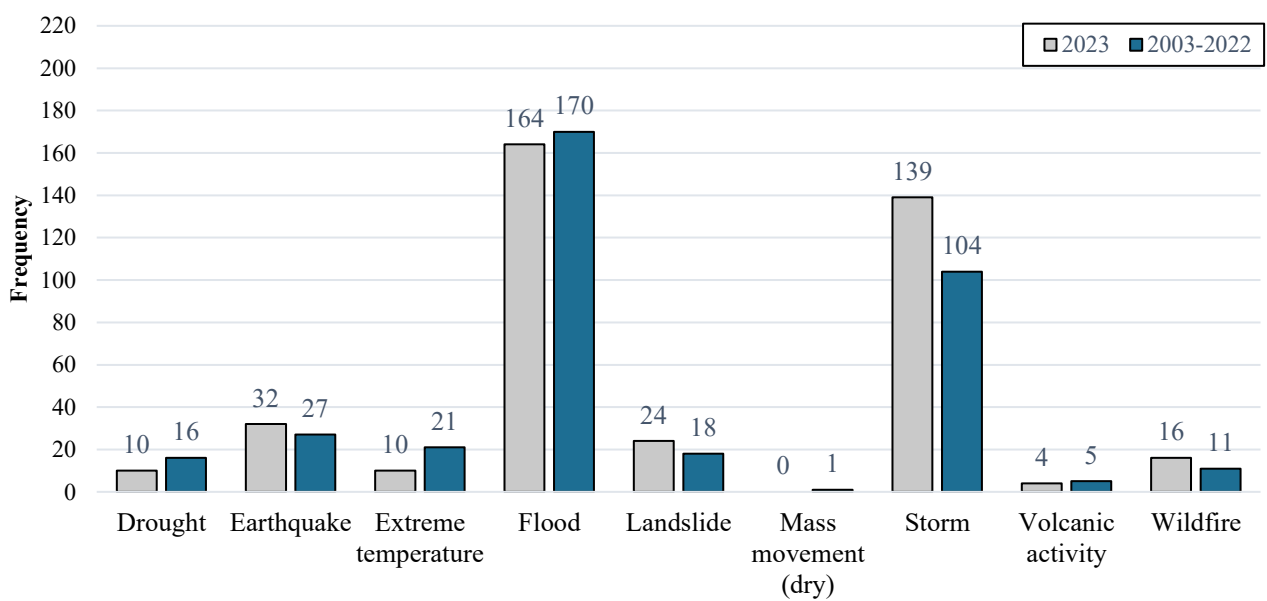

**Figure 1.** Comparison of disaster types occurrence -2023 versus Annual Average (2003-2022) [6]

Floods often result from multiple factors, such as heavy rainfall [7], rapid snowmelt [8, 9], and the rise of sea, lake [10, 11], river [12], and groundwater levels [13]. These contributing factors can cause an overflow of water that exceeds the capacity of natural or man-made drainage systems, leading to the inundation of land areas [14, 15]. Rapid unban expansion has also led to the displacement of impoverished populations into flood-prone areas, making them more vulnerable to flood hazards [16].

Flood risk assessment (FRA) can provide credible information to assist in the formulation of policies for managing floods, the distribution of resources, and the evaluation of the effectiveness of flood mitigation measures [17, 18]. In essence, flood risk assessment, which includes risk estimation and evaluation, serves as a critical component in the larger framework of "risk management". Risk estimation aims to determine the frequency and potential consequences of each selected risk scenario. Based on these estimates, risk evaluation integrates benefit-cost analysis and assesses the acceptability of risks to stakeholders, thereby identifying which risks necessitate immediate attention. These core steps underpin the subsequent phases of risk control and implementation, completing the risk management process [19, 20].

Flood risk assessment integrates factors like flood hazard, exposure, and vulnerability to determine the level of risk posed by flooding events [21, 22]. While relevant literature often share similar definitions and concepts for hazard and exposure, defining vulnerability and its various contributing factors has proven to be a significant challenge [23]. This is due in part to the complex and multifaceted nature of vulnerability, which can be influenced by a wide range of factors such as social, economic, environmental, and institutional conditions [1]. Certain articles focus on specific or restricted dimensions when it comes to risk assessment, such as the social dimension [2, 24, 25]. Additionally, these articles employ different perspectives to categorize hazard, vulnerability, and exposure criteria [26, 27]. As a result, indicators of hazard, vulnerability, and susceptibility or indicators of susceptibility and exposure may be used interchangeably.

Flood risk assessment is an emerging field. There has been a notable increase of publications on flood risk assessment over the last 25 years (1996–2020), more than one order of magnitude, and this trend has persisted over time without any notable declines [28]. Several studies in literature have conducted reviews on flood risk assessment. These studies often focus on bibliometric reviews, including tracking the number of publications per year and identifying leading countries in terms of publication output on the topic [28]. Other studies review current approaches to modeling the potential impacts of influential factors on flood risk, such as climate change, population growth, increasing urbanization, and infrastructure decay [29] or classifying the methods used in flood research [26]. Additionally, some reviews concentrate on specific components of flood risk, such as flood resilience [30]. To the authors' knowledge, no comprehensive scientific study currently exists that synthesize and integrates various flood risk frameworks proposed in the literature. Furthermore, there is a lack of an exhaustive compilation of established criteria for assessing risk components, systematically organized in a hierarchical structure. Therefore, to determine the most appropriate risk concept, assessment criteria, and the most suitable quantification method tailored to a specific study area and available data, a comprehensive literature review is essential, which facilitates well-informed decision-making.

Building upon the previous conceptual frameworks, we present a conceptual framework called Integrated Risk Linkages (IRL). Additionally, we develop three detailed hierarchical structures to outline the key criteria and sub-criteria within this framework. Recognizing that risk conceptual frameworks vary across studies, our aim is to provide a practical and adaptable model that integrates essential risk components. This aligns with recent studies emphasizing the importance of systematic criteria selection to improve risk assessment and decision-making processes [31, 32]. The proposed hierarchical structures incorporate physical, economic, social, political, environmental, infrastructural, and managerial factors. The paper is structured in three main steps: (1) a review of flood risk assessment conceptual frameworks in the literature is conducted to establish a foundational understanding, (2) an extensive set of factors influencing flood risk assessment is extracted from the literature and organized into a hierarchical structure encompassing three risk components—hazard, susceptibility (or exposure), and resilience, and (3) well-established methods for assessing flood risk are reviewed, with insights provided for future research directions.

**2. Methodology**

This review systematically identifies and evaluates key literature on flood risk assessment, with a particular emphasis on frameworks addressing hazards, susceptibility, and resilience, following the PRISMA methodology [33, 34]. Figure 2 illustrates the workflow for the literature review. Peer-reviewed English-language articles published between 2000 and 2023 were sourced using databases such as Google Scholar. In addition, authoritative and credible information was obtained from official resources, including the Federal Emergency Management Agency (https://www.fema.gov) and the World Conference on Disaster Risk Reduction (http://www.wcdrr.org). For the evaluation of resilience, the timeframe from 2010 to 2023 was specifically chosen based on expert recommendations and in alignment with the Sendai Framework for Disaster Risk Reduction (2015). This period was selected to capture recent advancements and evolving methodologies in disaster risk assessment, ensuring the study reflects contemporary developments in the field.

To ensure a systematic and robust selection process for the reviewed articles, specific inclusion and exclusion criteria were applied at both the title and abstract levels, as well as the full-article level. For inclusion based on titles and abstracts, we selected studies that addressed methods of flood risk assessment, such as MCDM, machine learning, remote sensing (RS), and GIS. Articles were also included if they focused on flood risk components, such as hazard, susceptibility, resilience, or broader concepts of risk components. At the full-article level, studies were further evaluated for their relevance and practical application. Specifically, articles were included if they applied indices related to hazard, vulnerability, susceptibility, resilience, or exposure. Conversely, exclusion criteria were applied to eliminate less relevant or redundant studies. Articles were rejected at the title and abstract level if they were review articles, book chapters, or published prior to a specific year. At the full-article level, studies were excluded if they contained duplicate indicators or flood risk methods, lacked full-text availability, or were written in non-English languages. This rigorous selection process ensured the inclusion of high-quality and diverse studies while maintaining the focus on advancing

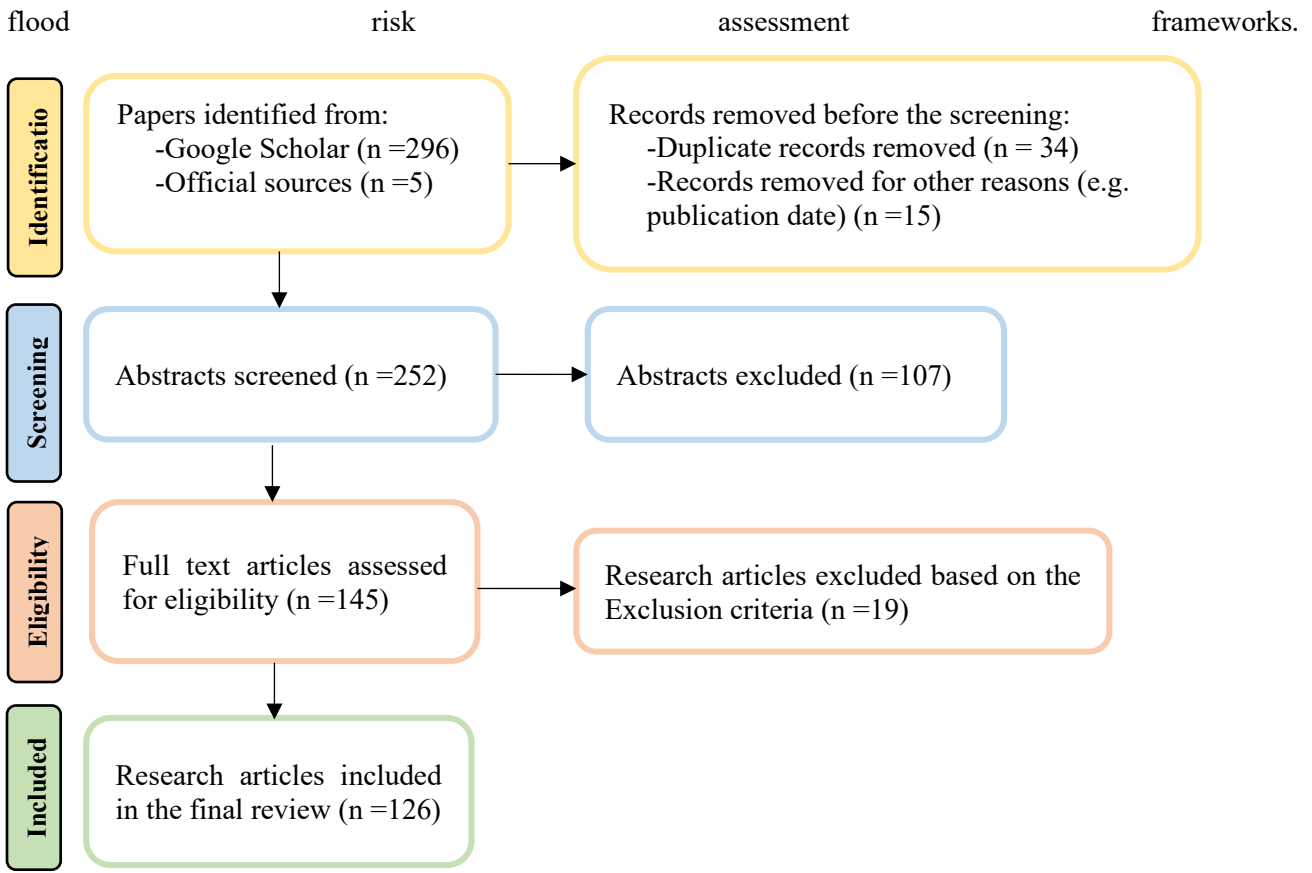

**Figure 2.** Illustration of the search strategy for literature review.

## 3. Flood Risk Conceptual Frameworks

Using hazard, exposure, and vulnerability as its receptors, Crichton's "risk triangle" methodology has been commonly used to estimate spatial patterns of disaster risks for flooding [35]. The triangle serves as a visual representation to demonstrate how changes in any one of these components influence overall risk [36]. According to this framework, surface water flooding constitutes a hazard; receptor vulnerability (refers to the inherent characteristics of urban communities); and exposure (is the geographic position of the urban environment), which can either exacerbate or mitigate the impact of the hazard [37]. This model's simplicity and clarity make it a foundational tool for understanding the interactions between risk components, laying the groundwork for more complex frameworks. The Pressure and Release (PAR) Model [38], highlights the progression of vulnerability in relation to hazards. Widely applied in emergency response contexts [39, 40], the model conceptualizes risk as a function of hazard and vulnerability, with a particular focus on factors contributing to susceptibility [41]. The traditional PAR model outlines three stages to explain how vulnerability evolves: root causes, dynamic pressures, and unsafe conditions. Each stage builds upon the previous one, increasing systemic pressure. When combined with the presence of a hazard, this progression leads to heightened disaster risk and, ultimately, to disaster itself [41, 42]. This structured approach underscores the interconnected nature of vulnerability and its role in disaster outcomes. More recently, the MOVE Framework (Method for the Improvement of Vulnerability in Europe), developed by [43], integrates hazard, exposure, vulnerability, and resilience into a comprehensive risk assessment approach. Designed to enhance the understanding of vulnerability's multi-faceted nature, the framework is particularly relevant for applications in disaster risk reduction and climate change adaptation [43]. This framework defines vulnerability through key causal factors, including exposure, susceptibility, and lack of resilience [44]. Furthermore, the MOVE Framework facilitates vulnerability assessments across diverse thematic dimensions, including physical, social, ecological, economic, cultural, and institutional aspects [45, 46]. Its holistic design supports nuanced evaluations of vulnerability and resilience within complex systems.

Hazard is defined as a potentially harmful phenomenon, substance, human activity, or condition that carries the potential to cause harm, damage, or pose a threat [47, 48]. Technically speaking, hazard can be understood as the expected value of annual inundation damage and losses (in monetary units) [1, 47]. More details on the criteria and sub-criteria can be found in section 5.1. Flood vulnerability is part of the flood risk concept [1, 2, 14, 16, 49, 50]. Vulnerability refers to the susceptibility of an area or population to the negative impacts of flooding [14]. Penning-Rowsell and Chatterton [51] defined susceptibility as the relative damageability of property and materials during floods or other hazardous events. Susceptibility is also defined as characteristics of exposed elements of a system, which influence the probabilities of damage at times of hazardous floods. Section 5.2 provides more information about the susceptibility criteria and sub-criteria. The United Nations Office for Disaster Risk Reduction (UNISDR) defines exposure as the presence of people, property, systems, or other elements within hazard zones, which are therefore vulnerable to potential losses [48, 49]. On the other hand, Balica [50] characterizes exposure as the values that exist in areas where floods may occur, including goods, infrastructure, cultural heritage, agricultural fields, and most importantly, people.

Inspired by existing conceptual frameworks and the established definitions of key risk components, we introduce the Integrated Risk Linkages (IRL) Framework, as depicted in Figure 3. This framework provides a simple yet comprehensive approach to risk assessment, emphasizing the interconnectedness of its components. It is commonly understood from literature that flood risk is a result of the interaction between hazard and vulnerability [1, 14, 16]. Therefore, the IRL framework initially illustrates the overlap between these concepts. From the managerial perspective, the exposure component is considered constant for the built environment and, therefore, is not easily manageable. In contrast, hazard and vulnerability are the primary contributors to risk variability and are more manageable components of risk (Figure 3a).

Vulnerability is a general term with various definitions (see supplementary Table S1). The exposure component has been defined either as a separate entity alongside vulnerability [52, 53] or as part of the vulnerability definition (Figure 3b). In the latter case, exposure is considered as physical vulnerability, which, when intersecting with intrinsic vulnerability (hereinafter referred to as susceptibility), results in overall vulnerability [14, 54]. This latter definition offers a clearer understanding of the concepts.

Over the past decade, the consensus in flood risk management has increasingly focused on concepts such as resilience, which is viewed as a positive factor that reduces overall risk [55–58]. Flood resilience can be expressed as the ability of a system or community to cope with, reduce or minimize flood damage [50]. In another definition, the concept of urban resilience is defined as enhancing the ability of cities to face adverse events and their inherent and adaptive capacity to respond and adapt, regardless of the type of disturbance they experience [59]. Resilience, as defined by the United Nations, refers to the ability of a system, community, or society that has been exposed to hazards to withstand, assimilate, adjust to, and recover from the consequences of a hazard in a timely and efficient manner, which includes preserving and restoring its essential structures and functions. This means that resilience can be thought of as the capacity to resist, recover from, or absorb shocks [48]. See section 5.3 for further information on the resilience criteria and sub-criteria.

Chen [60] suggested a new risk geometry for floods that incorporates community resilience into the risk triangle, positing that an increase in resilience diminishes the overall risk area. Accordingly, we can define resilience as a component that mitigates the vulnerability of communities exposed and susceptible to flooding hazards (Figure 3c). This conceptual framework can better clarify the roles that resilience plays in the overall framework. This conceptual framework provides a more nuanced perspective on the role of resilience within the broader flood risk landscape. It clarifies how resilience-building efforts targeting both coping and adaptive capacities (Figure 3d) can directly influence and reduce the risk faced by flood-prone communities, beyond just the hazard and exposure components traditionally considered in risk assessments.

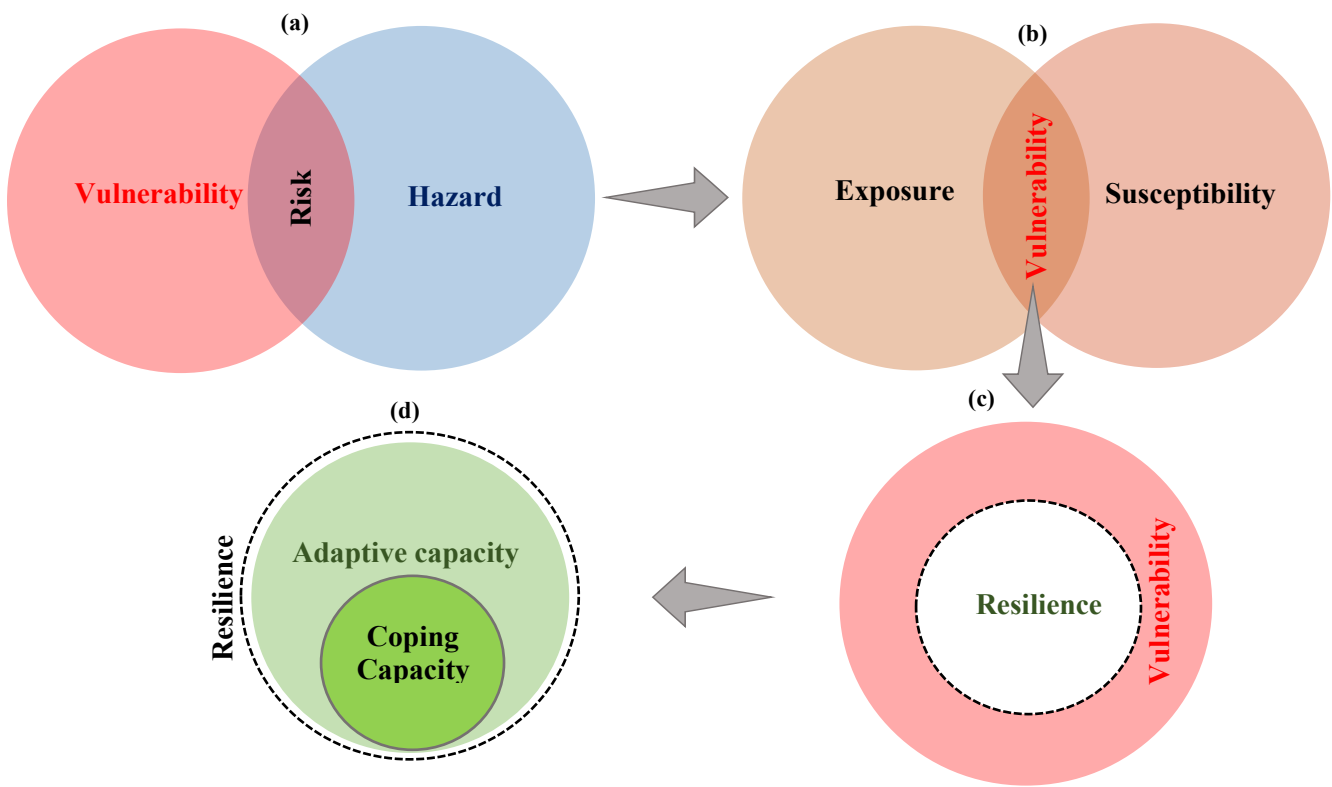

**Figure 3.** Integrated Risk Linkages (IRL) framework: **a)** Risk as a combination of vulnerability and hazard; **b)** intersection of exposure and susceptibility as the vulnerability; **c)** Contribution of resilience in reducing the vulnerability; **d)** the adaptive and coping capacities required to build the resilience.

## 4. Flood Risk Estimation

In order to assess the overall risk caused by floods, the Federal Emergency Management Agency (FEMA) [61] suggested Equation 1 in their 2018 guidelines:

$$R_{Total} = R_p + \sum_{k=1}^{3} R_{sk} = (H_p \times V_p) + \sum_{k=1}^{3}(H_{sk} \times V_{sk}) \quad k = 1.....3 \tag{1}$$

where $R_{Total}$ indicates the overall flood risk score, $R_p$ indicates the primary flood impact risk score. The secondary flood impact risk score is denoted as $R_{sk}$ where $k$ represents the type of impact (*i.e.*, $k$ =1 for technological hazard risk, $k$ =2 for geological hazard risk, and $k$ =3 for other impacts). In addition, the scores $H$ and $V$ represent the level of hazard and vulnerability[ 54, 55].

The primary impact of the flood is the damage resulting from the inundation [64]. Besides the direct consequences of flooding, cities can also experience effects like electrical injuries and drowning. Secondary impacts refer to the repercussions that arise as a result of the initial impacts. In certain cases, these secondary impacts can lead to greater destruction or losses compared to the primary impact [65, 66].

But as was stated in the section before, the risk assessment needs to account for community resilience. In this regard, as the vulnerability model is based on the principle that there is an inverse relationship between flood vulnerability and flood resilience ($Re$) [67], the model's parameters can be modified to reflect this negative correlation (Equation 2).

$$V = E + S - Re \tag{2}$$

The capacity of populations and assets to recover and withstand flood impacts becomes critical when they are extremely exposed.

## 5. Risk components and their criteria

### 5.1. Hazard

Flood hazard can be defined as the magnitude and location of a flood event that is expected to occur over a given period of time [14]. This definition emphasizes the spatial and temporal aspects of flood hazard. In the body of literature several criteria have been used to characterize the hazard such as average rainfall [15, 68–72], flood depth [2, 4, 24, 73–76], flood velocity [2, 4, 73], land slope [15, 69, 72, 75, 77, 78], drainage density [69, 72, 77–79], distance to channel (river) [3, 4, 15, 70, 77, 79], land use [3, 15, 77, 79, 80], flood duration [16, 68, 74, 76] and ground elevation (e.g., digital elevation model) [3, 4, 15, 15, 16, 75, 77, 78]. Among these indicators, slope, depth, elevation, and average rainfall have emerged as commonly used parameters for flood hazard assessment.

This study proposes a hierarchical structure to streamline the hazard assessment process (Figure 4). The framework begins with a critical decision point: "Is physically based flood mapping applicable??" If the answer is yes, the process transitions to analyzing flood characteristics, focusing on factors such as flood depth, velocity, and duration. Conversely, if the response is negative, the framework shifts to assessing area characteristics, including topographic and physiographic conditions. This dual-path approach ensures adaptability to diverse data availability and analytical requirements, enabling both detailed hydrodynamic modeling and flexible, criteria-driven hazard assessments.

When examining flood-prone areas, researchers often focus on two critical aspects: "topographic condition" and "physiographic condition". Topographic condition considers the physical features of the terrain. It encompasses aspects such as slope [2–4, 16, 78, 80–87], curvature [82–84, 88], topographic position index (TPI) [85, 87], flow accumulation [85, 89, 90], topographic wetness index (TWI) [3, 16, 81, 82, 84, 85] and sediment transport index (STI) [1, 91, 92]. Areas with lower elevations or steeper slopes are more susceptible to flooding. Additionally, the shape of the land influences water flow patterns and flood risk. Areas with poorly draining soils or dense vegetation with compacted soils beneath may experience higher flood risk due to reduced infiltration and increased runoff.

Understanding the physiographic features of a region, helps in identifying areas prone to river flooding and assessing flood risk levels [93, 94]. This subcategory includes factors like drainage density [69, 72, 75, 77–79, 81, 95, 96], time of concentration [97–99], distance to channel or river [3, 4, 15, 70, 77, 79], land cover [3, 15, 73, 79, 80, 82, 84–87, 96, 100–102]. The land surface is covered by natural or man-made structures. The criteria of land cover are analyzed using the factors of "permeability," "normalized difference vegetation index," and "land use". Land use classifications can be broadly categorized into vegetation-covered areas, water bodies, urban settlements, industrial and military zones, and other major infrastructures such as transportation networks and agricultural installations. Additionally, land use can be delineated based on surface coverage, including lithological features and soil types.

If flood mapping is conducted using physically-based models, the hazard component within the risk framework can be quantified through key factors such as flood depth [2–4, 24, 73–76], flood velocity [2, 4, 73] and flood duration [3, 16]. Flood depth typically represents the maximum water level reached during a flood event, providing insights into the extent of inundation. Flood velocity determines the erosive force and potential structural impacts, with higher velocities leading to increased erosion and scouring. Flood duration considers the length of time an area remains submerged, which is crucial for understanding prolonged water exposure and its effects on infrastructure, ecosystems, and communities.

It is important to note that these criteria are used to quantify hazards and may overlap with other risk components. For example, land cover is used in flood modeling but is also valuable for assessing vulnerability [103]. Additionally, some criteria, such as permeability and land use, may overlap and be used interchangeably. However, the primary focus of the current study is to gather these criteria from various literature sources and organize them to have a comprehensive hierarchical structure. Table S3 presents hazard criteria and its three sub-criteria levels along with their effects on flood risk and the corresponding references.

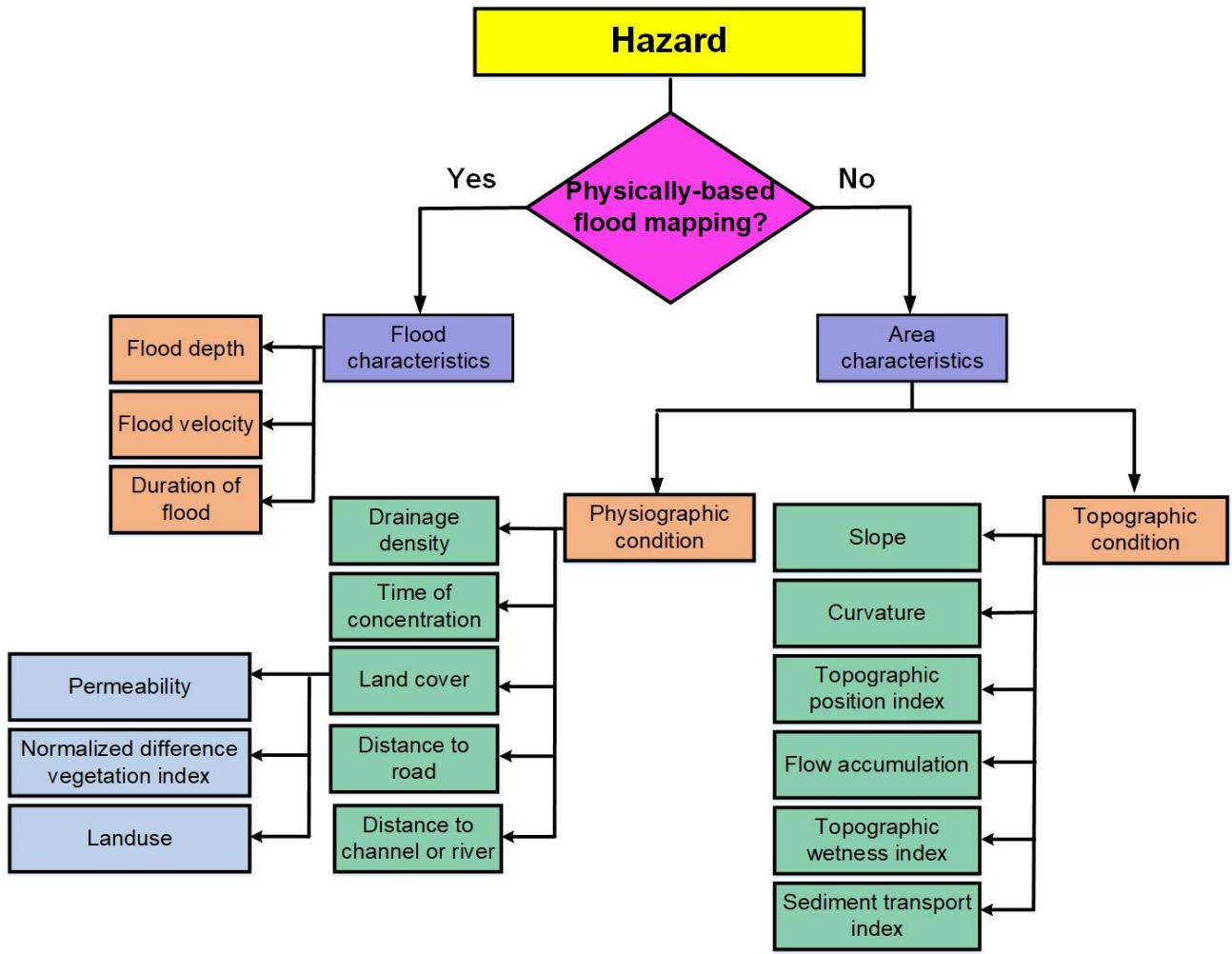

**Figure 4.** Hierarchical structure for quantifying flood hazard level

**5.2. Susceptibility**

The intersection of susceptibility and exposure is defined to be the vulnerability of a system to floods. However, in the literature there is not a clear distinction between vulnerability and susceptibility. Studies have shown that flood vulnerability can be assessed through various criteria such as literacy rate [104, 105], land use [16, 104], quality of building [27, 68], building density [27, 73], green area ratio (vegetation type) [76, 105], age of buildings [49, 68, 105], population density [15, 27, 68, 76, 79, 100, 104], drainage density [75], density of transportation networks [15, 70, 105] and age [49, 105]. As it seems, flood vulnerability is a multidimensional concept, involving various social, physical, and environmental components. Considering the journal papers centered around "susceptibility" term, various criteria influencing flood susceptibility have been highlighted such as vegetation Index [2, 80, 84, 85, 87, 101], DEM [81, 82, 84–87, 96, 101], Topographic position index [85, 87], flow accumulation [85], distance from river [82, 84, 87, 101], land use [82, 84–86, 96, 101], soil type [2, 80, 81, 84–86, 101], precipitation (rainfall) [81, 82, 84–86, 101], lithology (geology) [84, 86, 101], stream power index [82, 84, 101], curvature [82, 84, 101], river density [84, 87], building materials [80, 87], disabled people [16, 96], slope [16, 81, 82, 96], Topographic wetness index [16, 81, 82], drainage density [81, 82, 96] and average monthly household's income [16, 96]. The discussion of flood susceptibility often emphasizes social, economic, and demographic factors. This suggests that these factors represent the characteristics of an area that contribute to its vulnerability to flooding. Such factors include the condition of local infrastructure, the income status of populations at risk, as well as the geographical and topographical features of the affected region [16, 106].

Upon review of the literature, it appears that the criteria for exposure and susceptibility are also largely overlapping and might be interpreted differently. Exposure refers to the presence of people, assets, or infrastructure in areas prone to flooding, while susceptibility relates to the vulnerability of these elements to flood hazards [107–109]. Flood exposure criteria encompass various factors that contribute to the vulnerability of an area to flooding. These criteria include degree of soil erosion [73, 87, 104], popular density [24, 49, 73, 87, 102, 104, 105], maximum daily precipitation [49], days precipitation [49], surface runoff (mm/day) [49], elevation [76], slope [49, 76], impermeability [76], land use [73, 87, 100, 102], building area [4, 105], number of total cars [100], number of people [100], vegetation index [87], household size [16, 102], family type, damaged household in last floods events [16, 49] and building age [16]. These criteria indicate that demographic factors and regional characteristics, such as population density and property specifications, are commonly used in literature.

In certain cases, one criterion could be used to characterize both susceptibility and exposure. For instance, high population density increases the number of people exposed to flood damage, i.e., where flooding has not yet occurred but could potentially impact densely populated areas. Susceptibility indicates that individuals are already at risk of flood damage, with factors like population density exacerbating vulnerability. Thus, high population density exposes more individuals to flooding, amplifying the potential severity of damage. To address this issue, these two criteria are merged and organized as shown in Figure 5. Consequently, in evaluating the susceptibility and exposure of urban areas, managerial susceptibility, physical susceptibility, human susceptibility, and social susceptibility are considered as the main group of criteria. Moreover, Table S4 displays the collection of criteria, and three levels of sub-criteria utilized to assess flood susceptibility and exposure, as well as their impact on flood risk. As a part of crisis management, government authorities and organizations responsible for flood risk management engage in planning, monitoring, and implementing policies aimed at reducing or minimizing flood damage and losses. According to Figure 5b, the evaluation of managerial susceptibility includes the following sub-criteria: 1) weakness of education and information about flood risk management [110, 111], 2) insufficient financial assistance and offering loans [110, 112], 3) weakness in the watershed management planning [113] and 4) parallel responsibilities for flood management and planning [110, 113].

As shown in Figure 5c, physical susceptibility refers to the sensitivity of urban structures, infrastructure, and facilities to flooding. The key elements comprising a city, including residential buildings, transportation networks and urban facilities like fire stations, hospitals and schools play a crucial role in shaping its susceptibility to flood. The degree of physical susceptibility to floods is significantly influenced by the type, quality, and construction style of residential and public buildings, given that these structures are among the most critical facilities. The assessment of physical susceptibility for residential buildings, considers criteria such as the number of stories [16, 114, 115], worn-out texture [59] and urban density [15, 76]. In urban environments, the physical susceptibility to flooding is significantly influenced by the transportation network and road infrastructure, alongside residential structures and other urban facilities. The transportation network encompasses both the road network [15, 70, 105, 116, 117] and railroad network [68], catering to urban, suburban, and intercity travel requirements. Urban facilities encompass public institutions within cities, including critical infrastructure such as hospitals, fire stations, banks, schools, and malls and also urban infrastructure such as water, electricity, and gas facilities, all of which contribute to enhancing the quality of life for citizens. The evaluation of this criterion involves assessing urban infrastructure [110, 118], service centers [110, 118], and community infrastructure [87, 89, 110].

Another crucial aspect exposed to floods is the population residing in these urban areas. The criteria of human susceptibility examines factors such as population density [2, 15, 68, 70, 79, 104] and sensitive population [59] across different urban areas. Figure 5e outlines the sensitive population, which includes categories such as disabled population, female population, population with special needs (e.g. elderly individuals, and children who requiring assistance during emergencies) [59], and population affected by flood damage [49].

Urban communities, burdened by unfavorable social conditions, face heightened vulnerability and increased exposure to risk. To assess social susceptibility (Figure 5d), households having no means of transportation [16, 59, 119], households not having access to proper sanitation [16], Low-income households [16, 24], illiterate population [59, 70, 100], households without insurance coverage for accidents and health-related issues [16] and unemployed population in the area [59] are considered.

### 5.3. Resilience

In the body of literature, the definitions of resilience were found to frequently incorporate two concepts: (1) coping capacity and (2) adaptation capacity [56, 56, 120, 121]. As per the United Nations definition, coping capacity is the ability of individuals, organizations, and systems to confront and manage adverse emergency conditions using their available skills and resources [48]. Coping capacity also encompasses the ability to prevent permanent damage and ensure the provision of basic resources and services such as food, shelter, and search and rescue operations [120]. The United Nations defines adaptive capacity as the required adjustment in both natural and human systems in response to actual or anticipated climatic elements or their impacts, which helps to mitigate harm or take advantage of beneficial opportunities [48]. Essentially, post-disaster recovery and adaptation capacity relate to the necessary capabilities for rehabilitation and the ability to adapt through policy improvements [120].

In evaluating the resilience of urban areas, two key factors are examined: 1) management and planning and 2) available resources for treatment, rescue, supply basic need and repairing. Essentially, this means that in order to enhance resilience, the area should possess sufficient facilities to effectively respond, cope with, and adapt to flood events, as well as robust mechanisms for planning and managing flood-prone areas [51, 107, 109].

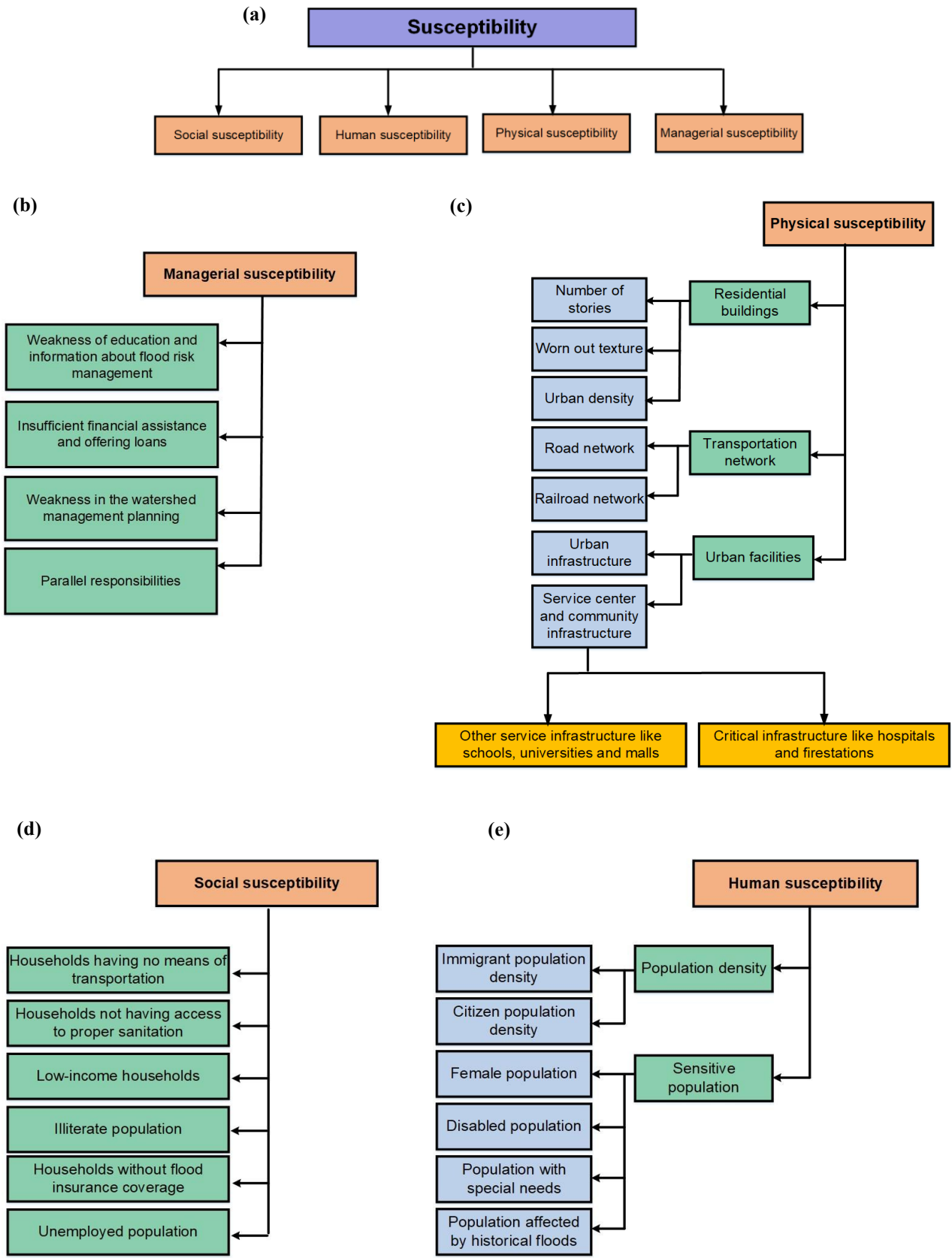

**Figure 5.** a) Hierarchical framework for flood susceptibility assessment, b) The criteria of managerial susceptibility in flood susceptibility hierarchical, c) The criteria of physical susceptibility in flood susceptibility hierarchical, d) The criteria of social susceptibility in flood susceptibility hierarchical, e) The criteria of human susceptibility in flood susceptibility hierarchical

These factors are assessed based on criteria and sub-criteria derived from literature reviews, insights from organizations like Crisis Management Organizations and Red Crescent Organizations, input from experts and specialists in the field, and the guidelines outlined in the Sendai framework for Disaster Risk Reduction [110].

Figure 6 presents a collection of criteria and sub-criteria to evaluate resilience in urban areas. The primary objective of integrated flood management and planning is to offer comprehensive solutions within the crisis organizations, with the aim of minimizing the loss of life and property resulting from floods, while also maximizing the effective utilization of their positive aspects [57, 122]. Based on the definition of flood risk management, this process is divided into two components: sufficiency of actions and sufficiency of planning. Figure 6b depicts the hierarchical structure of flood resilience management and planning, evaluating the effectiveness of crisis management organizations through key planning criteria: the level of task organization and task detailing [123], comprehensiveness of flood preparedness, response, and relief programs [124], frequency of review and updates [110], and leveraging past events and expert experiences [16, 120, 125]. These factors collectively determine the adequacy of planning in addressing flood risk management challenges, ensuring efficient response, optimal resource allocation, and effective implementation of risk reduction strategies. The sufficiency of actions pertains to the implementation of a set of measures aimed at mitigating or eliminating the risk posed by natural hazards to individuals and communities. These activities are categorized into two groups: structural plans (e.g., dredging canals, enhancing of surface water collection channels and drainage system, etc.) [116, 126] and non-structural plans (e.g., insurance map) [124].

Available resources for search, rescue, treatment, relief, reconstruction, and rehabilitation, as well as facilities for supplying basic needs, encompass various components such as vehicles, centers, trained members, and specialist manpower. To evaluate this factor, several sub-criteria are taken into account, namely search and rescue resource, medical and treatment resource, security resource, resource of basic needs supply, resource of urban infrastructure repairing and psychotherapy resource (psychological services). Figure 6c illustrates the hierarchical structure of available resources. Table S5 shows all levels of the identified criteria with their impacts on floods and the corresponding references

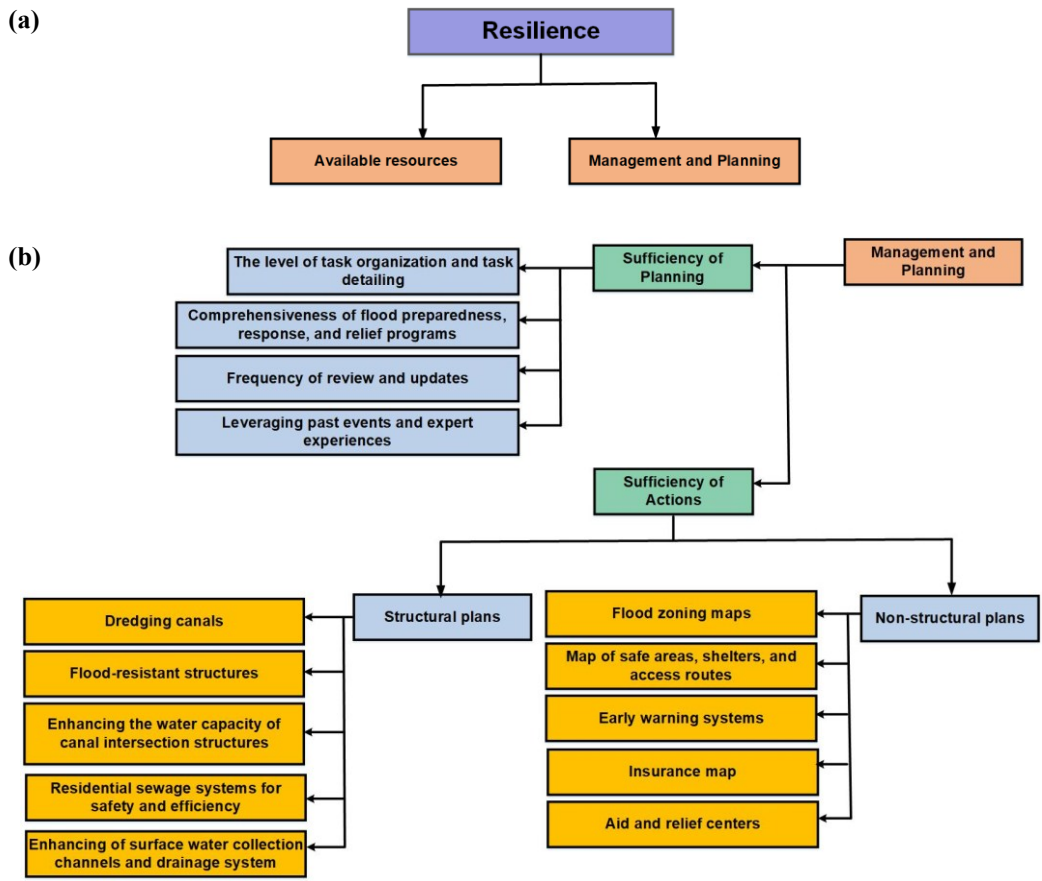

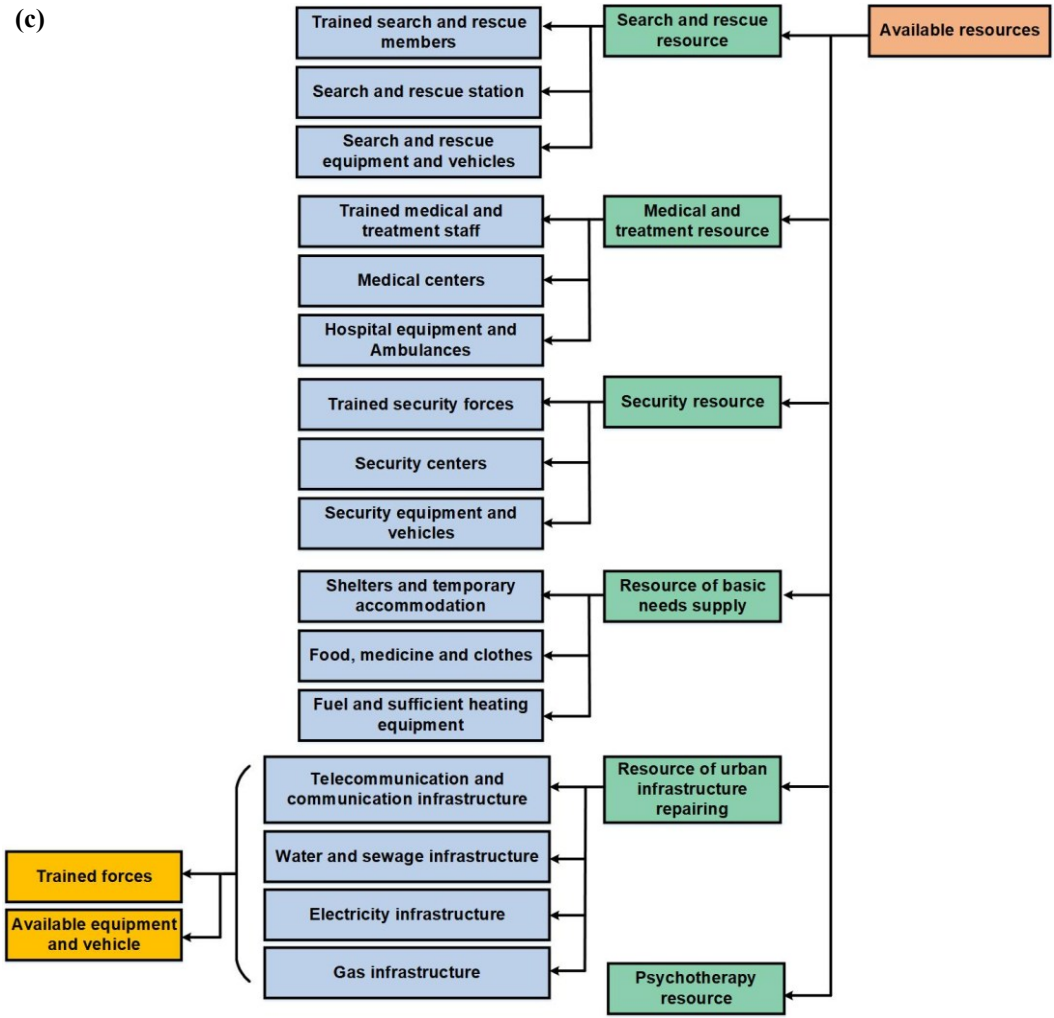

**Figure 6.** a) Conceptual hierarchical framework for flood resilience assessment, b) The hierarchical structure of management and planning in flood resilience, c) The hierarchical structure of available resources in the field of flood resilience.

## 6. Flood risk assessment Methods

Assessing flood risk can be categorized into three scales: micro, meso, and macro [127–129]. The macro-scale involves assessing flood risk at the national level, requiring national data. A meso-scale assessment focuses on evaluating flood risks within a province, catchment, or large city [49, 107]. The micro-scale is the smallest scale and pertains to a town or specific river stretch.

According to the findings of [22], there are five distinct kinds of flood assessment approaches. These include the historical disaster mathematical statistics method, the multi-criteria index system method, the remote sensing (RS), and GIS coupling method, the scenario simulation evaluation method, and the machine learning method. The suitability of each method depends on the assessment scale, study area's characteristics and data availability. This section reviews different urban flood risk assessment methods.

The mathematical statistics approach for historical disasters evaluates and forecasts hazards using data from past flood events. Since this approach depends heavily on historical records, it is not appropriate for situations in which data is scarce [27].

Multi-Criteria Analysis (MCA) as one of the qualitative approaches is often applied for complex decision-making problem in flood context by considering multiple factors. It involves integrating various criteria such as meteorological, geographical, economic, social, and environmental indicators to generate flood risk maps. There are many papers about MCA approaches on urban flood risk assessments. As the most frequent technique used in this context, AHP [68, 71, 76, 79, 87, 104, 130], Weighted Linear Combination (WLC) [79], Custom Weighted Average (OWA) [79], IAHP [4, 104], TOPSIS [4], FAHP [2, 68, 76], Shapley value and the Analytic Hierarchy Process method (SAHP) [105] have been extensively studied for their effectiveness in urban flood risk assessment. The integration of MCA with other techniques, such as deep learning, has shown promising results in identifying flood-susceptible areas [74]. Although MCA provides a comprehensive framework for assessing flood risk and developing effective mitigation strategies, it has limitations in determining subjective weights, as it primarily relies on experts' judgments for decision-making. To gain a deeper understanding of the relative significance of the main criteria and sub-criteria developed in this study, expert opinions were utilized to assign weights to each criterion using three AHP techniques.

The method, which combines RS and GIS, involves using remote sensing technology to gather data about the disaster area, including water area, inundation duration, and disaster-bearing bodies that could be affected by disasters. This data is then entered into GIS software for spatial analysis. When studying large-scale flood disasters, the use of RS technology makes it quicker and easier to collect information on the research area's flood risk. Nevertheless, remote sensing data often fails to adequately capture the flood process during small floods because of their brief duration, and their temporal and geographical resolution is severely limited [120]. One important area of future research will be the integrated application of GIS and multi-source fusion data for flood risk estimation. For the evaluation of flood risk, scenario-based simulation analysis involves quantifying flood levels and assessing the extent of inundation in different scenarios. In addition to providing data support for disaster risk transfer, this approach may intuitively and accurately present the spatial distribution features of urban flood disaster risk, which can serve as a point of reference for managers making decisions about risk management and disaster prevention and mitigation [131]. This approach is a quantitative method that commonly used to predict inundation risk in a small region, but flood disaster usually happened in a regional scale [29]. A variety of models and software tools are utilized for this analysis, including mathematical and hydrodynamic models [37, 76], and GIS-based models [132], and HEC suite tools (e.g., HEC-RAS [100, 133, 134], HEC-GeoRAS, HEC-FDA [133, 134], and HEC-WAT [135]). To assess the risk associated with each scenario, the first step involves quantifying the extent of damage caused by floods with different return periods [73].

Data mining methods have been increasingly used in flood risk assessment to improve accuracy and efficiency of the analysis [129, 136, 137]. These methods have been applied in various aspects of flood risk assessment, including the development of mitigation measures, emergency response preparation, and flood recovery planning [137].Utilizing data mining techniques in flood risk assessment involves extracting specific patterns and relationships from substantial hydraulic and hydrological data across one or multiple extensive databases. A significant challenge with these approaches is their reliance on precise hydrological and hydraulic data, making them unsuitable for regions with limited data availability. This is particularly problematic in developing countries, where accessing data is a major hurdle [15, 138]. The interested readers are referred to [22] for additional details on the comparison of flood risk assessment techniques.

Researchers and practitioners often combine multiple methods (e.g. studies like [139] and [77]) to enhance the overall flood risk assessment [140]. But it is necessary to consider key factors when selecting a flood risk assessment method. One of these factors is spatial scale. Some methods like scenario-based inundation analysis and GIS or RS techniques are better suited for micro analyses, while others work well at larger scales [22]. Readers are encouraged to read [128] for more details on the characteristics of assessments at various scales. Another factor is the availability and quality of data required for each method. This includes topographic data, land use/land cover data, hydrological data, historical flood records, and climate data. Statistical methods, GIS or RS techniques and data mining methods need reliable data to assess flood risk [141]. Methods relying on extensive data may be challenging in areas with limited data availability. Statistical methods are useful for historical analysis, while machine learning models can predict future events. The complexity of each method should be assessed in order to balance the trade-off between accuracy and resource requirements [121, 142]. It should be noted that, all methods have inherent uncertainties which should be recognized [22, 83]. Before selecting method for flood risk assessment, the sensitivity of results to input parameters and assumptions should be evaluated [144, 145].

## 7. Discussion

Flood risk assessment forms a vital cornerstone for urban flood management, guiding decisions on disaster prevention, mitigation, and relief efforts. Given the multifaceted nature of flood disasters—driven by complex and dynamic processes—several challenges persist in achieving accurate and actionable insights. A critical examination of the existing body of literature highlights these challenges while informing future research. The question arises whether current flood risk assessment research offers sufficiently accurate and reliable results for practical applications. From this perspective, the IRL framework (Figure 3) can be utilized. Its flexibility allows researchers and practitioners to tailor the framework and utilize either specific components or the entire hierarchical structure, depending on the availability of data and their research goals. This approach aligns with studies emphasizing the need for integrated frameworks to improve risk assessment outcomes [146–148].

The configuration of the framework's components is arbitrary and often dictated by the availability and quality of data. In cases where reliable vulnerability data is scarce or insufficient, a simplified approach—combining hazard data with an exposure indicator—can effectively identify flood hotspots. This streamlined method provides actionable insights for practitioners, enabling them to prioritize risk mitigation projects efficiently [149, 150]. Such an approach is particularly valuable in resource-limited settings where rapid decision-making is essential. However, it may not fully capture the complexities of urban flood vulnerability. If the depth-damage function linked to component vulnerability is expressed in monetary terms, risk can be derived solely from the interaction of vulnerability and hazard, rendering the exposure component unnecessary [36].

Flood risk management is increasingly adopting a multi-dimensional perspective, incorporating social, economic, environmental, cultural, and policy factors to better understand and address urban flood vulnerability [151]. For more sophisticated applications, such as advanced urban planning, disaster risk reduction programs, or financial analyses in the insurance sector, a more comprehensive integration of the vulnerability component becomes critical [152–155]. However, integrating these diverse dimensions introduces challenges related to the availability, quality, and resolution of vulnerability data [156–158]. The heterogeneous nature of such data complicates its integration into comprehensive frameworks. Moreover, simply increasing the number of input data does not necessarily translate into greater accuracy in flood risk assessments; on the contrary, it may lead to higher uncertainty if the data quality or resolution is not suitable [121, 159, 160].

As noted by [36], a practical approach to address confusion in risk-related terminology is to adopt general definitions and prioritize assessments over detailed theoretical distinctions. The comprehensive list of indicators identified for hazard, susceptibility and resilience ensures a holistic assessment and facilitates the integration of data from multiple sources, capturing the inherent complexity of the risk assessment process. Basically, it promotes interdisciplinary collaboration by uniting experts across fields to tackle the diverse challenges of flood risk management. More importantly, the framework allows for the selection of criteria based on data availability, ensuring its adaptability to different contexts. For practical applications, we recommend establishing a framework with the best available high-quality data to quantify flood risk as precisely as possible. Subsequently, alternative frameworks can be developed using all available data, including lower-quality datasets. This approach enables decision-makers to assess the uncertainty associated with risk quantification due to the lack of precise or high-resolution data.

Once the framework is established, its application can be tailored to the quantitative or qualitative nature of the available data. This flexibility enables the use of methodologies such as multi-criteria decision-making (MCDM) analysis, modeling, or a combination of these approaches, as illustrated in Figure 4. In particular, advancements in artificial intelligence and big data analytics offer transformative opportunities for flood risk assessment. These technologies enable rapid flood mapping [161], providing near real-time insights that are invaluable for emergency response and mitigation planning. Furthermore, they support refined multi-dimensional flood risk assessments [162], which integrate diverse datasets to capture the complex interactions between hazards, exposure, and vulnerability. Future research should focus on improving data collection methods, enhancing the resolution and quality of socio-economic datasets, and developing tools that balance computational demands with accessibility. Additionally, expanding the framework to incorporate emerging technologies more seamlessly and exploring innovative approaches for uncertainty quantification in data-limited contexts will be essential, ultimately supporting more effective disaster risk management and urban resilience strategies.

## 8. Conclusion

Risk assessment plays a vital role in mitigating the impact of floods and has broad applications in disaster preparedness and flood management. This study presented an integrated risk linkages (IRL) framework for the assessment of flood-related risks, developed through a thorough review of the literature. The proposed framework integrates a wide array of pertinent criteria, encompassing the full spectrum from hazard estimation to disaster management strategies. This is particularly important given the complex nature of flood hazards and their potential aftermath. The intricacies of assessing flood risk in urban environments were acknowledged in this paper, with a focus on primary impacts of flooding. In this context, two distinct components were identified: hazard and vulnerability. Vulnerability was further divided into three components: exposure, susceptibility, and resilience. Exposure and susceptibility negatively impact vulnerability, while resilience has a positive impact. Based on the defined concepts and interrelationships of the key vulnerability components, the study has developed three comprehensive, hierarchical assessment structures. These structures capture the multidimensional aspects of flood-related hazards, susceptibility, and resilience, including 21, 30, and 40 distinct criteria/sub-criteria, respectively. Moreover, we reviewed commonly used methods for evaluating flood risk in the literature, including scenario-based analysis, data mining techniques, and multi-criteria analysis.

The application of the framework can be tailored based on the analysis scale and selected assessment methods. For in-depth hazard analysis at smaller scales, the use of high-resolution hydraulic and hydrologic modeling is recommended to capture the nuances of the flood dynamics. However, quantifying the vulnerability aspect of the risk assessment poses a greater challenge due to the large number of important but qualitative criteria, which can introduce substantial uncertainty. To address this challenge, a combined approach leveraging simulation models, data-driven models, and multi-criteria analysis techniques could provide a promising future research avenue. This proposed methodology would enable a more holistic and robust evaluation of flood vulnerability, accounting for both the quantifiable and qualitative factors that contribute to overall risk. The IRL framework presented in this study offers essential insights for navigating the complexities of flood risk management, serving as a valuable reference for researchers, policymakers, and practitioners.

**Acknowledgements** We would like to gratefully acknowledge Dr. Majid Salari from the Industrial Engineering Department at Ferdowsi University of Mashhad for his continuous support throughout the research and his invaluable comments.

**Funding** This research received no specific grant from public, commercial, or not-for-profit funding agencies.

**Author contributions** N.T. and M.F. collaborated on the initial draft and conceptualization of the study. M.F. and B.R. reviewed and revised the manuscript for intellectual content. N.T. and M.F. prepared the literature review tables and Figures. All authors contributed to later versions of the manuscript. All authors read and approved the final manuscript.

**Data availability** The authors confirm that the data supporting the findings of this study are available within the article and its supplementary materials.

**Declaration** The authors of this manuscript affirm that all contributors have reviewed and provided their consent regarding the content presented herein.

**Competing interests** The authors declare that they have no competing interests.

**References**

1. Tsakiris, G.: Flood risk assessment: concepts, modelling, applications. Nat. Hazards Earth Syst. Sci. 14, 1361–1369 (2014). https://doi.org/10.5194/nhess-14-1361-2014
2. Zou, Q., Zhou, J., Zhou, C., Song, L., Guo, J.: Comprehensive flood risk assessment based on set pair analysis-variable fuzzy sets model and fuzzy AHP. Stoch. Environ. Res. Risk Assess. 27, 525–546 (2013). https://doi.org/10.1007/s00477-012-0598-5
3. Wang, Z., Lai, C., Chen, X., Yang, B., Zhao, S., Bai, X.: Flood hazard risk assessment model based on random forest. J. Hydrol. 527, 1130–1141 (2015). https://doi.org/10.1016/j.jhydrol.2015.06.008
4. Xu, H., Ma, C., Lian, J., Xu, K., Chaima, E.: Urban flooding risk assessment based on an integrated k-means cluster algorithm and improved entropy weight method in the region of Haikou, China. J. Hydrol. 563, 975–986 (2018). https://doi.org/10.1016/j.jhydrol.2018.06.060
5. Cea, L., Costabile, P.: Flood risk in urban areas: modelling, management and adaptation to climate change. A review. Hydrology. 9, 50 (2022). https://doi.org/10.3390/hydrology9030050
6. Centre for Research on the Epidemiology of Disasters (CRED): Disaster year in review 2020: Global trends and perspectives. Cred Crunch. 62, (2021). https://legacy.cred.be/
7. Ariyani, D., Perdinan, Purwanto, M.Y.J., Sunarti, E., Juniati, A.T., Ibrahim, M.: Contributed Indicators to Fluvial Flood Along River Basin in Urban Area of Indonesia. Geogr. Environ. Sustain. 15, 102–114 (2023). https://doi.org/10.24057/2071-9388-2022-084
8. Angelakis, A.N., Capodaglio, A.G., Valipour, M., Krasilnikoff, J., Ahmed, A.T., Mandi, L., Tzanakakis, V.A., Baba, A., Kumar, R., Zheng, X., Min, Z., Han, M., Turay, B., Bilgiç, E., Dercas, N.: Evolution of Floods: From Ancient Times to the Present Times (ca 7600 BC to the Present) and the Future. Land. 12, 1211 (2023). https://doi.org/10.3390/land12061211
9. Goodarzi, M.R., Sabaghzadeh, M., Niazkar, M.: Evaluation of snowmelt impacts on flood flows based on remote sensing using SRM model. Water. 15, 1650 (2023). https://doi.org/10.3390/w15091650
10. Befus, K., Barnard, P.L., Hoover, D.J., Finzi Hart, J., Voss, C.I.: Increasing threat of coastal groundwater hazards from sea-level rise in California. Nat. Clim. Change. 10, 946–952 (2020). https://doi.org/10.1038/s41558-020-0874-1
11. Yu, X., Luo, L., Hu, P., Tu, X., Chen, X., Wei, J.: Impacts of sea-level rise on groundwater inundation and river floods under changing climate. J. Hydrol. 614, 128554 (2022). https://doi.org/10.1016/j.jhydrol.2022.128554
12. Nigatu, G.T., Abebe, B.A., Grum, B., Kebedew, M.G., Semane, E.M.: Investigation of Flood incidence causes and mitigation: Case study of Ribb river, northwestern Ethiopia. Nat. Hazards Res. 3, 408–419 (2023). https://doi.org/10.1016/j.nhres.2023.04.009
13. Bosserelle, A.L., Morgan, L.K., Hughes, M.W.: Groundwater rise and associated flooding in coastal settlements due to sea-level rise: a review of processes and methods. Earths Future. 10, (2022). https://doi.org/10.1029/2021EF002580
14. Olesen, L., Löwe, R., Arnbjerg-Nielsen, K.: Flood damage assessment–Literature review and recommended procedure. (2017). https://watersensitivecities.org.au/content/flood-damage-assessment-literature-review-recommended-procedure/
15. Darabi, H., Choubin, B., Rahmati, O., Haghighi, A.T., Pradhan, B., Kløve, B.: Urban flood risk mapping using the GARP and QUEST models: A comparative study of machine learning techniques. J. Hydrol. 569, 142–154 (2019). https://doi.org/10.1016/j.jhydrol.2018.12.002
16. Rana, I.A., Routray, J.K.: Integrated methodology for flood risk assessment and application in urban communities of Pakistan. Nat. Hazards. 91, 239–266 (2018). https://doi.org/10.1007/s11069-017-3124-8
17. Hall, J.W., Dawson, R., Sayers, P., Rosu, C., Chatterton, J., Deakin, R.: A methodology for national-scale flood risk assessment. In: Proceedings of the Institution of Civil Engineers. pp. 235–247. Thomas Telford Ltd (2003)
18. Ekmekcioğlu, Ö., Koc, K., Özger, M.: District based flood risk assessment in Istanbul using fuzzy analytical hierarchy process. Stoch. Environ. Res. Risk Assess. 35, 617–637 (2021). https://doi.org/10.1007/s00477-020-01924-8
19. Canadian Standards Association: Risk management: Guidelines for decision-makers (CAN/CSA-Q850-97). Canadian Standards Association, Rexdale (1997)

20. Simonović, S.P.: Floods in a changing climate: Risk management. Cambridge University Press, New York (2012)
21. Zhang, K., Shalehy, M.H., Ezaz, G.T., Chakraborty, A., Mohib, K.M., Liu, L.: An integrated flood risk assessment approach based on coupled hydrological-hydraulic modeling and bottom-up hazard vulnerability analysis. Environ. Model. Softw. 148, 105279 (2022). https://doi.org/10.1016/j.envsoft.2021.105279
22. Li, C., Sun, N., Lu, Y., Guo, B., Wang, Y., Sun, X., Yao, Y.: Review on urban flood risk assessment. Sustainability. (2023). https://doi.org/10.3390/su15010765
23. Anwana, E.O., Owojori, O.M.: Analysis of flooding vulnerability in informal settlements literature: mapping and research agenda. Soc. Sci. 12, 40 (2023). https://doi.org/10.3390/socsci12010040
24. Koks, E.E., Jongman, B., Husby, T.G., Botzen, W.J.: Combining hazard, exposure and social vulnerability to provide lessons for flood risk management. Environ. Sci. Policy. 47, 42–52 (2015). https://doi.org/10.1016/j.envsci.2014.10.013
25. Rufat, S., Tate, E., Burton, C.G., Maroof, A.S.: Social vulnerability to floods: Review of case studies and implications for measurement. Int. J. Disaster Risk Reduct. 14, 470–486 (2015). https://doi.org/10.1016/j.ijdrr.2015.09.013
26. Diaconu, D.C., Costache, R., Popa, M.C.: An overview of flood risk analysis methods. Water. 13, 474 (2021). https://doi.org/10.3390/w13040474
27. Peng, J., Zhang, J.: Urban flooding risk assessment based on GIS-game theory combination weight: A case study of Zhengzhou City. Int. J. Disaster Risk Reduct. 77, 103080 (2022). https://doi.org/10.1016/j.ijdrr.2022.103080
28. Díez-Herrero, A., Garrote, J.: Flood risk analysis and assessment, applications and uncertainties: A bibliometric review. Water. 12, 2050 (2020). https://doi.org/10.3390/w12072050
29. Salman, A.M., Li, Y.: Flood risk assessment, future trend modeling, and risk communication: a review of ongoing research. Nat. Hazards Rev. 19, 04018011 (2018). https://doi.org/10.1061/(ASCE)NH.1527-6996.0000294
30. Herath, H., Wijesekera, N.: A State-of-the-Art Review of Flood Risk Assessment in Urban Area. Presented at the IOP Conference Series: Earth and Environmental Science (2019)
31. De Ruiter, M.C., Ward, P.J., Daniell, J.E., Aerts, J.C.: A comparison of flood and earthquake vulnerability assessment indicators. Nat. Hazards Earth Syst. Sci. 17, 1231–1251 (2017)
32. Gallina, V., Torresan, S., Critto, A., Sperotto, A., Glade, T., Marcomini, A.: A review of multi-risk methodologies for natural hazards: Consequences and challenges for a climate change impact assessment. J. Environ. Manage. 168, 123–132 (2016)
33. Dhawale, R., Wallace, C.S., Pietroniro, A.: Assessing the multidimensional nature of flood and drought vulnerability index: A systematic review of literature. Int. J. Disaster Risk Reduct. 104764 (2024). https://doi.org/10.1016/j.ijdrr.2024.104764
34. Liberati, A., Altman, D.G., Tetzlaff, J., Mulrow, C., Gøtzsche, P.C., Ioannidis, J.P., Clarke, M., Devereaux, P.J., Kleijnen, J., Moher, D.: The PRISMA statement for reporting systematic reviews and meta-analyses of studies that evaluate health care interventions: explanation and elaboration. Ann. Intern. Med. 151, W-65 (2009). https://doi.org/10.1371/journal.pmed.1000100
35. Crichton, D.: The implications of climate change for the insurance industry: an update and outlook to 2020. Building Research Establishment, Watford, England (2001)
36. Wolf, S.: Vulnerability and risk: comparing assessment approaches. Nat. Hazards. 61, 1099–1113 (2012). https://doi.org/10.1007/s11069-011-9968-4
37. Li, C., Cheng, X., Li, N., Du, X., Yu, Q., Kan, G.: A framework for flood risk analysis and benefit assessment of flood control measures in urban areas. Int. J. Environ. Res. Public. Health. 13, 787 (2016). https://doi.org/10.3390/ijerph13080787
38. Blaikie, P., Cannon, T., Davis, I., Wisner, B.: At risk: natural hazards, people's vulnerability and disasters. Routledge (2014)
39. Mustafa, D.: Structural causes of vulnerability to flood hazard in Pakistan. Econ. Geogr. 74, 289–305 (1998). https://doi.org/10.1111/j.1944-8287.1998.tb00117.x
40. Rauken, T., Kelman, I.: River flood vulnerability in Norway through the pressure and release model. J. Flood Risk Manag. 3, 314–322 (2010). https://doi.org/10.1111/j.1753-318X.2010.01080.x
41. Hammer, C.C., Brainard, J., Innes, A., Hunter, P.R.: (Re-) conceptualising vulnerability as a part of risk in global health emergency response: updating the pressure and release model for global health emergencies. Emerg. Themes Epidemiol. 16, 1–8 (2019)
42. Awal, M.: Vulnerability to disaster: Pressure and release model for climate change hazards in Bangladesh. Int. J. Environ. Monit. Prot. 2, 15–21 (2015). http://www.openscienceonline.com/journal/archive2?journalId=714&paperId=1663
43. Birkmann, J., Cardona, O.D., Carreño, M.L., Barbat, A.H., Pelling, M., Schneiderbauer, S., Kienberger, S., Keiler, M., Alexander, D., Zeil, P.: Framing vulnerability, risk and societal responses: the MOVE framework. Nat. Hazards. 67, 193–211 (2013). https://doi.org/10.1007/s11069-013-0558-5
44. Welle, T., Depietri, Y., Angignard, M., Birkmann, J., Renaud, F., Greiving, S.: Vulnerability assessment to heat waves, floods, and earthquakes Using the MOVE framework: Test case Cologne, Germany. In: Assessment of vulnerability to natural hazards. pp. 91–124. Elsevier (2014)
45. Kablan, M.K.A., Dongo, K., Coulibaly, M.: Assessment of social vulnerability to flood in urban Côte d'Ivoire using the MOVE framework. Water. 9, 292 (2017). https://doi.org/10.3390/w9040292
46. Hamidi, A.R., Wang, J., Guo, S., Zeng, Z.: Flood vulnerability assessment using MOVE framework: A case study of the northern part of district Peshawar, Pakistan. Nat. Hazards. 101, 385–408 (2020). https://doi.org/10.1007/s11069-020-03878-0

47. Pistrika, A., Tsakiris, G.: Flood Risk Assessment: A Methodological Framework. In: EWRA Symposium "Water Resources Management: New Approaches and Technologies." pp. 13–22. , Chania (Greece) (2007)
48. UNISDR, T.: Basic terms of disaster risk reduction. U. N. Int. Strategy Disaster Reduct. UNISDR Geneva. (2004)
49. Jun, K.-S., Chung, E.-S., Kim, Y.-G., Kim, Y.: A fuzzy multi-criteria approach to flood risk vulnerability in South Korea by considering climate change impacts. Expert Syst. Appl. 40, 1003–1013 (2013). https://doi.org/10.1016/j.eswa.2012.08.013
50. Balica, S.F.: Development and application of flood vulnerability indices for various spatial scales. U. N. Educ. Sci. Cult. Organ. Delft Netherlands—IHE Inst. Water Educ. (2007)
51. Penning-Rowsell, E.C., Chatterton, J.B.: The benefits of flood alleviation. A manual of assessment techniques. Gower Technical Press, Aldershot (1977)
52. Modica, M., Reggiani, A., Nijkamp, P.: Vulnerability, resilience and exposure: Methodological aspects. Adv. Spat. Econ. Model. Disaster Impacts. 295–324 (2019). https://doi.org/10.1007/978-3-030-16237-5_12
53. Antwi-Agyei, P., Dougill, A.J., Fraser, E.D., Stringer, L.C.: Characterising the nature of household vulnerability to climate variability: Empirical evidence from two regions of Ghana. Environ. Dev. Sustain. 15, 903–926 (2013). https://doi.org/10.1007/s10668-012-9418-9
54. Xofi, M., Domingues, J.C., Santos, P.P., Pereira, S., Oliveira, S.C., Reis, E., Zêzere, J.L., Garcia, R.A., Lourenço, P.B., Ferreira, T.M.: Exposure and physical vulnerability indicators to assess seismic risk in urban areas: a step towards a multi-hazard risk analysis. Geomat. Nat. Hazards Risk. 13, 1154–1177 (2022). https://doi.org/10.1080/19475705.2022.2068457
55. Wang, L., Cui, S., Li, Y., Huang, H., Manandhar, B., Nitivattananon, V., Fang, X., Huang, W.: A review of the flood management: from flood control to flood resilience. Heliyon. 8, (2022). https://doi.org/10.1016/j.heliyon.2022.e11763
56. Bertilsson, L., Wiklund, K., de Moura Tebaldi, I., Rezende, O.M., Veról, A.P., Miguez, M.G.: Urban flood resilience–A multi-criteria index to integrate flood resilience into urban planning. J. Hydrol. 573, 970–982 (2019). https://doi.org/10.1016/j.jhydrol.2018.06.052
57. Waghwala, R.K., Agnihotri, P.: Flood risk assessment and resilience strategies for flood risk management: A case study of Surat City. Int. J. Disaster Risk Reduct. 40, 101155 (2019). https://doi.org/10.1016/j.ijdrr.2019.101155
58. Islam, M., Fereshtehpour, M., Najafi, M., Khaliq, M., Khan, A., Sushama, L., Nguyen, V., Elshorbagy, A., Roy, R., Wilson, A.: Climate-resilience of dams and levees in Canada: a review. Discov. Appl. Sci. 6, 174 (2024). https://doi.org/10.1007/s42452-024-05814-4
59. Moghadas, M., Asadzadeh, A., Vafeidis, A., Fekete, A., Kötter, T.: A multi-criteria approach for assessing urban flood resilience in Tehran, Iran. Int. J. Disaster Risk Reduct. 35, 101069 (2019). https://doi.org/10.1016/j.ijdrr.2019.101069
60. Chen, I.: New conceptual framework for flood risk assessment in Sheffield, UK. Geogr. Res. 59, 465–482 (2021). https://doi.org/10.1111/1745-5871.12478
61. FEMA: Hazus 6.1 Release Notes. 36 (2023). https://www.fema.gov/flood-maps/tools-resources/flood-map-products/hazus/release-notes
62. Kreibich, H., Di Baldassarre, G., Vorogushyn, S., Aerts, J.C., Apel, H., Aronica, G.T., Arnbjerg-Nielsen, K., Bouwer, L.M., Bubeck, P., Caloiero, T.: Adaptation to flood risk: Results of international paired flood event studies. Earths Future. 5, 953–965 (2017). https://doi.org/10.1002/2017EF000606
63. Petrucci, O.: Factors leading to the occurrence of flood fatalities: a systematic review of research papers published between 2010 and 2020. Nat. Hazards Earth Syst. Sci. 22, 71–83 (2022). https://doi.org/10.5194/nhess-22-71-2022
64. Mei, C., Liu, J., Wang, H., Li, Z., Yang, Z., Shao, W., Ding, X., Weng, B., Yu, Y., Yan, D.: Urban flood inundation and damage assessment based on numerical simulations of design rainstorms with different characteristics. Sci. China Technol. Sci. 63, 2292–2304 (2020). https://doi.org/10.1007/s11431-019-1523-2
65. Doocy, S., Daniels, A., Murray, S., Kirsch, T.D.: The human impact of floods: a historical review of events 1980-2009 and systematic literature review. PLoS Curr. 5, (2013). https://doi.org/10.1371/currents.dis.f4deb457904936b07c09daa98ee8171a
66. Tillihal, S.B., Shukla, A.K.: Flood disaster hazards: a state-of-the-art review of causes, impacts, and monitoring. Adv. Water Resour. Plan. Sustain. 77–95 (2023). https://doi.org/10.1007/978-981-99-3660-1_5
67. Sun, R., Shi, S., Reheman, Y., Li, S.: Measurement of urban flood resilience using a quantitative model based on the correlation of vulnerability and resilience. Int. J. Disaster Risk Reduct. 82, 103344 (2022). https://doi.org/10.1016/j.ijdrr.2022.103344
68. Lyu, H.-M., Shen, S.-L., Zhou, A.-N., Zhou, W.-H.: Flood risk assessment of metro systems in a subsiding environment using the interval FAHP-FCA approach. Sustain. Cities Soc. 50, 101682 (2019). https://doi.org/10.1016/j.scs.2019.101682
69. Chen, W., Li, Y., Xue, W., Shahabi, H., Li, S., Hong, H., Wang, X., Bian, H., Zhang, S., Pradhan, B.: Modeling flood susceptibility using data-driven approaches of naïve bayes tree, alternating decision tree, and random forest methods. Sci. Total Environ. 701, 134979 (2020). https://doi.org/10.1016/j.scitotenv.2019.134979
70. Mishra, K., Sinha, R.: Flood risk assessment in the Kosi megafan using multi-criteria decision analysis: A hydro-geomorphic approach. Geomorphology. 350, 106861 (2020). https://doi.org/10.1016/j.geomorph.2019.106861
71. Chen, Y., Liu, R., Barrett, D., Gao, L., Zhou, M., Renzullo, L., Emelyanova, I.: A spatial assessment framework for evaluating flood risk under extreme climates. Sci. Total Environ. 538, 512–523 (2015). https://doi.org/10.1016/j.scitotenv.2015.08.094
72. Joo, H., Choi, C., Kim, J., Kim, D., Kim, S., Kim, H.S.: A Bayesian network-based integrated for flood risk assessment (InFRA). Sustainability. 11, 3733 (2019). https://doi.org/10.3390/su11133733

73. Foudi, S., Osés-Eraso, N., Tamayo, I.: Integrated spatial flood risk assessment: The case of Zaragoza. Land Use Policy. 42, 278–292 (2015). https://doi.org/10.1016/j.landusepol.2014.08.002
74. Pham, B.T., Luu, C., Van Dao, D., Van Phong, T., Nguyen, H.D., Van Le, H., von Meding, J., Prakash, I.: Flood risk assessment using deep learning integrated with multi-criteria decision analysis. Knowl.-Based Syst. 219, 106899 (2021). https://doi.org/10.1016/j.knosys.2021.106899
75. Malekian, A., Azarnivand, A.: Application of integrated Shannon's entropy and VIKOR techniques in prioritization of flood risk in the Shemshak watershed, Iran. Water Resour. Manag. 30, 409–425 (2016). https://doi.org/10.1007/s11269-015-1169-6
76. Cai, T., Li, X., Ding, X., Wang, J., Zhan, J.: Flood risk assessment based on hydrodynamic model and fuzzy comprehensive evaluation with GIS technique. Int. J. Disaster Risk Reduct. 35, 101077 (2019). https://doi.org/10.1016/j.ijdrr.2019.101077
77. Taromideh, F., Fazloula, R., Choubin, B., Emadi, A., Berndtsson, R.: Urban flood-risk assessment: Integration of decision-making and machine learning. Sustainability. 14, 4483 (2022). https://doi.org/10.3390/su14084483
78. Li, G., Wu, X., Han, J.-C., Li, B., Huang, Y., Wang, Y.: Flood risk assessment by using an interpretative structural modeling based Bayesian network approach (ISM-BN): An urban-level analysis of Shenzhen, China. J. Environ. Manage. 329, 117040 (2023). https://doi.org/10.1016/j.jenvman.2022.117040
79. Kandilioti, G., Makropoulos, C.: Preliminary flood risk assessment: the case of Athens. Nat. Hazards. 61, 441–468 (2012). https://doi.org/10.1007/s11069-011-9930-5
80. Zeleňáková, M., Gaňová, L., Purcz, P., Satrapa, L.: Methodology of flood risk assessment from flash floods based on hazard and vulnerability of the river basin. Nat. Hazards. 79, 2055–2071 (2015). https://doi.org/10.1007/s11069-015-1945-x
81. Zhao, G., Pang, B., Xu, Z., Peng, D., Zuo, D.: Urban flood susceptibility assessment based on convolutional neural networks. J. Hydrol. 590, 125235 (2020). https://doi.org/10.1016/j.jhydrol.2020.125235
82. Khosravi, K., Pham, B.T., Chapi, K., Shirzadi, A., Shahabi, H., Revhaug, I., Prakash, I., Bui, D.T.: A comparative assessment of decision trees algorithms for flash flood susceptibility modeling at Haraz watershed, northern Iran. Sci. Total Environ. 627, 744–755 (2018). https://doi.org/10.1016/j.scitotenv.2018.01.266
83. Arabameri, A., Pourghasemi, H.R.: Spatial modeling of gully erosion using linear and quadratic discriminant analyses in GIS and R. In: Spatial modeling in GIS and R for earth and environmental sciences. pp. 299–321. Elsevier (2019)
84. Bui, D.T., Tsangaratos, P., Ngo, P.-T.T., Pham, T.D., Pham, B.T.: Flash flood susceptibility modeling using an optimized fuzzy rule based feature selection technique and tree based ensemble methods. Sci. Total Environ. 668, 1038–1054 (2019). https://doi.org/10.1016/j.scitotenv.2019.02.422
85. Kanani-Sadat, Y., Arabsheibani, R., Karimipour, F., Nasseri, M.: A new approach to flood susceptibility assessment in data-scarce and ungauged regions based on GIS-based hybrid multi criteria decision-making method. J. Hydrol. 572, 17–31 (2019). https://doi.org/10.1016/j.jhydrol.2019.02.034
86. Nkeki, F.N., Bello, E.I., Agbaje, I.G.: Flood risk mapping and urban infrastructural susceptibility assessment using a GIS and analytic hierarchical raster fusion approach in the Ona River Basin, Nigeria. Int. J. Disaster Risk Reduct. 77, 103097 (2022). https://doi.org/10.1016/j.ijdrr.2022.103097
87. Lin, L., Wu, Z., Liang, Q.: Urban flood susceptibility analysis using a GIS-based multi-criteria analysis framework. Nat. Hazards. 97, 455–475 (2019). https://doi.org/10.1007/s11069-019-03615-2
88. Khosravi, K., Pourghasemi, H.R., Chapi, K., Bahri, M.: Flash flood susceptibility analysis and its mapping using different bivariate models in Iran: a comparison between Shannon's entropy, statistical index, and weighting factor models. Environ. Monit. Assess. 188, 1–21 (2016). https://doi.org/10.1007/s10661-016-5665-9
89. Tarboton, D.G., Bras, R.L., Rodriguez-Iturbe, I.: On the extraction of channel networks from digital elevation data. Hydrol. Process. 5, 81–100 (1991). https://doi.org/10.1002/hyp.3360050107
90. Jenson, S.K., Domingue, J.O.: Extracting topographic structure from digital elevation data for geographic information system analysis. Photogramm. Eng. Remote Sens. 54, 1593–1600 (1988)
91. Strahler, A., Strahler, A.N.: Introducing physical geography. Wiley, New York (1994)
92. Tehrany, M.S., Jones, S., Shabani, F.: Identifying the essential flood conditioning factors for flood prone area mapping using machine learning techniques. Catena. 175, 174–192 (2019). https://doi.org/10.1016/j.catena.2018.12.011
93. Ali, K., Bajracharya, R.M., Koirala, H.L.: A review of flood risk assessment. Int. J. Environ. Agric. Biotechnol. 1, 238636 (2016). https://doi.org/10.22161/ijeab/1.4.62
94. Anees, M.T., Bakar, A.F.B.A., San, L.H., Abdullah, K., Nordin, M.N.M., Ab Rahman, N.N.N., Ishak, M.I.S., Kadir, M.O.A.: Flood vulnerability, risk, and susceptibility assessment: Flood risk management. In: Decision Support Methods for Assessing Flood Risk and Vulnerability. pp. 1–27. IGI Global (2020)
95. Saidi, S., Ghattassi, A., Anselme, B., Bouri, S.: GIS based multi-criteria analysis for flood risk assessment: Case of manouba essijoumi basin, NE Tunisia. Presented at the Advances in Remote Sensing and Geo Informatics Applications: Proceedings of the 1st Springer Conference of the Arabian Journal of Geosciences (CAJG-1), Tunisia 2018 (2019)
96. Cikmaz, B.A., Yildirim, E., Demir, I.: Flood susceptibility mapping using fuzzy analytical hierarchy process for Cedar Rapids, Iowa. Int. J. River Basin Manag. 1–13 (2023). https://doi.org/10.1080/15715124.2023.2216936
97. Thompson, D., Fang, X., Cleverland, T.: Literature Review on Time Parameters, for Hydrographs. Dep. Civ. Eng. Lamar Univercity. (2004)

98. McCuen, R.H.: Hydrologic analysis and design. . Prentice-Hall, Upper Saddler River, NJ (2005)
99. Lo, S.: Glossary of hydrology. Water Resources Publications, Littleton, CO,USA (1992)
100. Gain, A.K., Mojtahed, V., Biscaro, C., Balbi, S., Giupponi, C.: An integrated approach of flood risk assessment in the eastern part of Dhaka City. Nat. Hazards. 79, 1499–1530 (2015). https://doi.org/10.1007/s11069-015-1911-7
101. Arabameri, A., Rezaei, K., Cerdà, A., Conoscenti, C., Kalantari, Z.: A comparison of statistical methods and multi-criteria decision making to map flood hazard susceptibility in Northern Iran. Sci. Total Environ. 660, 443–458 (2019). https://doi.org/10.1016/j.scitotenv.2019.01.021
102. Gacu, J.G., Monjardin, C.E.F., Senoro, D.B., Tan, F.J.: Flood risk assessment using GIS-based analytical hierarchy process in the municipality of Odiongan, Romblon, Philippines. Appl. Sci. 12, 9456 (2022). https://doi.org/10.3390/app12199456
103. Liu, J., Shi, Z.: Quantifying land-use change impacts on the dynamic evolution of flood vulnerability. Land Use Policy. 65, 198–210 (2017). https://doi.org/10.1016/j.landusepol.2017.04.012
104. Lin, K., Chen, H., Xu, C.-Y., Yan, P., Lan, T., Liu, Z., Dong, C.: Assessment of flash flood risk based on improved analytic hierarchy process method and integrated maximum likelihood clustering algorithm. J. Hydrol. 584, 124696 (2020). https://doi.org/10.1016/j.jhydrol.2020.124696
105. Yu, J., Zou, L., Xia, J., Chen, X., Wang, F., Zuo, L.: A multi-dimensional framework for improving flood risk assessment: Application in the Han River Basin, China. J. Hydrol. Reg. Stud. 47, 101434 (2023). https://doi.org/10.1016/j.ejrh.2023.101434
106. Kaya, C.M., Derin, L.: Parameters and methods used in flood susceptibility mapping: a review. J. Water Clim. Change. 14, 1935–1960 (2023). https://doi.org/10.2166/wcc.2023.035
107. Tate, E., Rahman, M.A., Emrich, C.T., Sampson, C.C.: Flood exposure and social vulnerability in the United States. Nat. Hazards. 106, 435–457 (2021). https://doi.org/10.1007/s11069-020-04470-2
108. Aydin, M.C., Sevgi Birincioğlu, E.: Flood risk analysis using gis-based analytical hierarchy process: a case study of Bitlis Province. Appl. Water Sci. 12, 122 (2022). https://doi.org/10.1007/s13201-022-01655-x
109. Bonasia, R., Cea, L., Fratino, U.: Flood Susceptibility and Risk Maps as a Crucial Tool to Face the Hydrological Extremes in Developing Countries: Technical and Governance Aspects Linked by a Participatory Approach. Front. Earth Sci. 10, 838172 (2022). https://doi.org/10.3389/feart.2022.838172
110. Center, U. (United N.I.S. for D.R.: Sendai framework for disaster risk reduction 2015–2030. U. N. Off. Disaster Risk Reduct. Geneva Switz. (2015). http://www.wcdrr.org/uploads/Sendai_Framework_for_ Disaster_Risk_Reduction_2015-2030.pdf
111. Tyszka, T., Zielonka, P.: Large risks with low probabilities: Perceptions and willingness to take preventive measures against flooding. IWA Publishing (2017)
112. FEMA: Disaster Financial Management Guide: Guidance for State, Local, Tribal & Territorial Partners, Guidance for State, Local, Tribal and Territorial Partners. Exec. Agency Publ. HS 5.108:F 49, (2020)
113. Brooks, K.N., Ffolliott, P.F., Magner, J.A.: Hydrology and the Management of Watersheds. John Wiley & Sons, USA (2013)
114. Vogel, K., Riggelsen, C., Scherbaum, F., Schröter, K., Kreibich, H., Merz, B.: Challenges for Bayesian network learning in a flood damage assessment application. Presented at the 11th International Conference on Structural Safety and Reliability (2013)
115. Deniz, D., Arneson, E.E., Liel, A.B., Dashti, S., Javernick-Will, A.N.: Flood loss models for residential buildings, based on the 2013 Colorado floods. Nat. Hazards. 85, 977–1003 (2017). https://doi.org/10.1007/s11069-016-2615-3
116. Kalantari, Z., Ferreira, C.S.S., Koutsouris, A.J., Ahlmer, A.-K., Cerdà, A., Destouni, G.: Assessing flood probability for transportation infrastructure based on catchment characteristics, sediment connectivity and remotely sensed soil moisture. Sci. Total Environ. 661, 393–406 (2019). https://doi.org/10.1016/j.scitotenv.2019.01.009
117. Diakakis, M., Boufidis, N., Grau, J.M.S., Andreadakis, E., Stamos, I.: A systematic assessment of the effects of extreme flash floods on transportation infrastructure and circulation: The example of the 2017 Mandra flood. Int. J. Disaster Risk Reduct. 47, 101542 (2020). https://doi.org/10.1016/j.ijdrr.2020.101542
118. Pearson, L., Pelling, M.: The UN Sendai Framework for Disaster Risk Reduction 2015–2030: Negotiation Process and Prospects for Science and Practice. J. Extreme Events. 02, 1571001 (2015). https://doi.org/10.1142/S2345737615710013
119. Flanagan, B.E., Gregory, E.W., Hallisey, E.J., Heitgerd, J.L., Lewis, B.: A social vulnerability index for disaster management. J. Homel. Secur. Emerg. Manag. 8, (2011). https://doi.org/10.2202/1547-7355.1792
120. Li, Z., Zhang, X., Ma, Y., Feng, C., Hajiyev, A.: A multi-criteria decision making method for urban flood resilience evaluation with hybrid uncertainties. Int. J. Disaster Risk Reduct. 36, (2019). https://doi.org/10.1016/j.ijdrr.2019.101140
121. Fereshtehpour, M., Karamouz, M.: DEM resolution effects on coastal flood vulnerability assessment: Deterministic and probabilistic approach. Water Resour. Res. 54, 4965–4982 (2018). https://doi.org/10.1029/2017WR022318
122. Mai, T., Mushtaq, S., Reardon-Smith, K., Webb, P., Stone, R., Kath, J., An-Vo, D.-A.: Defining flood risk management strategies: A systems approach. Int. J. Disaster Risk Reduct. 47, (2020). https://doi.org/10.1016/j.ijdrr.2020.101550
123. HE, B., LIANG, G., ZHOU, H.: Workflow-based integration management of flood control task. Adv. Water Sci. 18, 900–906 (2007)
124. Wang, M., Fu, X., Zhang, D., Chen, F., Su, J., Zhou, S., Li, J., Zhong, Y., Tan, S.K.: Urban Flooding Risk Assessment in the Rural-Urban Fringe Based on a Bayesian Classifier. Sustainability. 15, 5740 (2023). https://doi.org/10.3390/su15075740
125. Tyler, J., Sadiq, A.-A., Noonan, D.S.: A review of the community flood risk management literature in the USA: Lessons for improving community resilience to floods. Nat. Hazards. 96, 1223–1248 (2019). https://doi.org/10.1007/s11069-019-03606-3

126. FEMA: Reducing damage from localized flooding a guide for communities, https://www.fema.gov/pdf/fima/FEMA511-complete.pdf, (2005)
127. ZHOU, Z.-Y., CHEN, W.-L., LI, X., YU, W.: Urban Micro-scale Flood Risk Evaluation Based on Scenario Simulation. DEStech Trans. Eng. Technol. Res. (2016)
128. De Moel, H., Jongman, B., Kreibich, H., Merz, B., Penning-Rowsell, E., Ward, P.J.: Flood risk assessments at different spatial scales. Mitig. Adapt. Strateg. Glob. Change. 20, 865–890 (2015). https://doi.org/10.1007/s11027-015-9654-z
129. Țîncu, R., Zêzere, J.L., Crăciun, I., Lazăr, G., Lazăr, I.: Quantitative micro-scale flood risk assessment in a section of the Trotuș River, Romania. Land Use Policy. 95, 103881 (2020). https://doi.org/10.1016/j.landusepol.2019.02.040
130. Chen, Y.-R., Yeh, C.-H., Yu, B.: Integrated application of the analytic hierarchy process and the geographic information system for flood risk assessment and flood plain management in Taiwan. Nat. Hazards. 59, 1261–1276 (2011). https://doi.org/10.1007/s11069-011-9831-7
131. Renn, O., Walker, K.: Global risk governance. In: Concept and practice using the IRGC framework. pp. 77–86. Springer, Netherland (2008)
132. Shaddoud, M.A., Costache, R., Kotaridis, I., Fereshtehpour, M., Kuriqi, A.: Flash flood prioritization assessment using morphometric analysis in the coastal region of the Eastern Mediterranean. DYSONA-Appl. Sci. 6, 172–185 (2025). https://doi.org/10.30493/das.2024.480819
133. Mohammadi, S., Nazariha, M., Mehrdadi, N.: Flood damage estimate (quantity), using HEC-FDA model. Case study: the Neka river. Procedia Eng. 70, 1173–1182 (2014)
134. Mahmood, S., Rahman, A., Shaw, R.: Spatial appraisal of flood risk assessment and evaluation using integrated hydro-probabilistic approach in Panjkora River Basin, Pakistan. Environ. Monit. Assess. 191, 1–15 (2019)
135. Dunn, C., Baker, P., Fleming, M.: Flood risk management with HEC-WAT and the FRA compute option. Presented at the E3S Web of Conferences (2016)
136. Ait Naceur, H., Igmoullan, B., Namous, M.: Machine learning-based optimization of flood susceptibility mapping in semi-arid zone. DYSONA-Appl. Sci. 6, 145–159 (2025). https://doi.org/10.30493/das.2024.483211
137. Alvan Romero, N., Cigna, F., Tapete, D.: ERS-1/2 and Sentinel-1 SAR data mining for flood hazard and risk assessment in Lima, Peru. Appl. Sci. 10, 6598 (2020)
138. Rahmati, O., Pourghasemi, H.R.: Identification of critical flood prone areas in data-scarce and ungauged regions: a comparison of three data mining models. Water Resour. Manag. 31, 1473–1487 (2017)
139. Lu, Y., Qin, X., Xie, Y.: An integrated statistical and data-driven framework for supporting flood risk analysis under climate change. J. Hydrol. 533, 28–39 (2016). https://doi.org/10.1016/j.jhydrol.2015.11.041
140. Malakeel, G.S., Abdu Rahiman, K., Vishnudas, S.: Flood Risk Assessment Methods—A Review. Curr. Trends Civ. Eng. Sel. Proc. ICRACE 2020. 197–208 (2021). https://doi.org/10.1007/978-981-15-8151-9_19
141. Ahmad, S., Jia, H., Ashraf, A., Yin, D., Chen, Z., Ahmed, R., Israr, M.: A Novel GIS-SWMM-ABM Approach for Flood Risk Assessment in Data-Scarce Urban Drainage Systems. Water. 16, 1464 (2024). https://doi.org/10.3390/w16111464
142. Awah, L.S., Belle, J.A., Nyam, Y.S., Orimoloye, I.R.: A Systematic Analysis of Systems Approach and Flood Risk Management Research: Trends, Gaps, and Opportunities. Int. J. Disaster Risk Sci. 1–13 (2024). https://doi.org/10.1007/s13753-024-00544-y
143. Nasiri, H., Mohd Yusof, M.J., Mohammad Ali, T.A.: An overview to flood vulnerability assessment methods. Sustain. Water Resour. Manag. 2, 331–336 (2016). https://doi.org/10.1007/s40899-016-0051-x
144. Sharma, S.K., Kwak, Y.-J., Kumar, R., Sarma, B.: Analysis of hydrological sensitivity for flood risk assessment. ISPRS Int. J. Geo-Inf. 7, 51 (2018). https://doi.org/10.3390/ijgi7020051
145. Koks, E.E., Bočkarjova, M., de Moel, H., Aerts, J.C.: Integrated direct and indirect flood risk modeling: development and sensitivity analysis. Risk Anal. 35, 882–900 (2015). https://doi.org/10.1111/risa.12300
146. Aven, T.: On some recent definitions and analysis frameworks for risk, vulnerability, and resilience. Risk Anal. Int. J. 31, 515–522 (2011)
147. Birkmann, J.: Risk and vulnerability indicators at different scales: Applicability, usefulness and policy implications. Environ. Hazards. 7, 20–31 (2007)
148. Fekete, A., Damm, M., Birkmann, J.: Scales as a challenge for vulnerability assessment. Nat. Hazards. 55, 729–747 (2010)
149. Aggarwal, A.: Exposure, hazard and risk mapping during a flood event using open source geospatial technology. Geomat. Nat. Hazards Risk. 7, 1426–1441 (2016). https://doi.org/10.1080/19475705.2015.1069408
150. Jalayer, F., De Risi, R., De Paola, F., Giugni, M., Manfredi, G., Gasparini, P., Topa, M.E., Yonas, N., Yeshitela, K., Nebebe, A.: Probabilistic GIS-based method for delineation of urban flooding risk hotspots. Nat. Hazards. 73, 975–1001 (2014). https://doi.org/10.1007/s11069-014-1119-2
151. Depietri, Y.: The social–ecological dimension of vulnerability and risk to natural hazards. Sustain. Sci. 15, 587–604 (2020). https://doi.org/10.1007/s11625-019-00710-y
152. Fedeski, M., Gwilliam, J.: Urban sustainability in the presence of flood and geological hazards: The development of a GIS-based vulnerability and risk assessment methodology. Landsc. Urban Plan. 83, 50–61 (2007). https://doi.org/10.1016/j.landurbplan.2007.05.012

153. Highfield, W.E., Peacock, W.G., Van Zandt, S.: Mitigation planning: Why hazard exposure, structural vulnerability, and social vulnerability matter. J. Plan. Educ. Res. 34, 287–300 (2014). https://doi.org/10.1177/0739456X1453182
154. Wing, O.E., Pinter, N., Bates, P.D., Kousky, C.: New insights into US flood vulnerability revealed from flood insurance big data. Nat. Commun. 11, 1444 (2020). https://doi.org/10.1038/s41467-020-15264-2
155. Bernardini, G., Ferreira, T.M., Julià, P.B., Eudave, R.R., Quagliarini, E.: Assessing the spatiotemporal impact of users' exposure and vulnerability to flood risk in urban built environments. Sustain. Cities Soc. 100, 105043 (2024). https://doi.org/10.1016/j.scs.2023.105043
156. Apel, H., Aronica, G.T., Kreibich, H., Thieken, A.H.: Flood risk analyses—how detailed do we need to be? Nat. Hazards. 49, 79–98 (2009). https://doi.org/10.1007/s11069-008-9277-8
157. Karamouz, M., Fereshtehpour, M., Ahmadvand, F., Zahmatkesh, Z.: Coastal flood damage estimator: An alternative to FEMA's HAZUS platform. J. Irrig. Drain. Eng. 142, 04016016 (2016). https://doi.org/10.1061/(ASCE)IR.1943-4774.0001017
158. Membele, G.M., Naidu, M., Mutanga, O.: Examining flood vulnerability mapping approaches in developing countries: A scoping review. Int. J. Disaster Risk Reduct. 69, 102766 (2022). https://doi.org/10.1016/j.ijdrr.2021.102766
159. Da Silva, L.B.L., Alencar, M.H., de Almeida, A.T.: Exploring global sensitivity analysis on a risk-based MCDM/A model to support urban adaptation policies against floods. Int. J. Disaster Risk Reduct. 73, 102898 (2022). https://doi.org/10.1016/j.ijdrr.2022.102898
160. Rehman, S., Sahana, M., Hong, H., Sajjad, H., Ahmed, B.B.: A systematic review on approaches and methods used for flood vulnerability assessment: framework for future research. Nat. Hazards. 96, 975–998 (2019). https://doi.org/10.1007/s11069-018-03567-z
161. Fereshtehpour, M., Esmaeilzadeh, M., Alipour, R.S., Burian, S.J.: Impacts of DEM type and resolution on deep learning-based flood inundation mapping. Earth Sci. Inform. 1–21 (2024)
162. Paswan, N.G., Ray, L.K.: Intelligent Solutions for Flood Management: Integrating Artificial Intelligence and Machine Learning. In: Big Data, Artificial Intelligence, and Data Analytics in Climate Change Research: For Sustainable Development Goals. pp. 43–55. Springer (2024)